\documentclass[twocolumn]{aastex62}
\usepackage{amsmath}
\usepackage{threeparttable}
\usepackage{CJK}

\def\Iuv{I_{\rm uv}}
\def\Iir{I_{\rm ir}}
\def\kapuv{\kappa_{\rm uv}}
\def\kapir{\kappa_{\rm ir}}

\shorttitle{Dusty Cloud Acceleration}
\shortauthors{}

\begin{document}
\begin{CJK*}{UTF8}{gbsn}

\title{Dusty Cloud Acceleration with Multiband Radiation}
\author[0000-0003-2868-489X]{Xiaoshan Huang (黄小珊)}
\affiliation{Department of Astronomy, University of Virginia, Charlottesville, VA 22904, USA}
\author[0000-0001-7488-4468]{Shane W. Davis}
\affiliation{Department of Astronomy, University of Virginia, Charlottesville, VA 22904, USA}
\author{Dong Zhang}
\affiliation{Department of Astronomy, University of Virginia, Charlottesville, VA 22904, USA}
\affiliation{Department of Astronomy, University of Michigan,
Ann Arbor, MI 48109, USA}

\begin{abstract}

  We perform two-dimensional and three-dimensional simulations of cold, dense clouds, which are accelerated by radiation pressure on dust relative to a hot, diffuse background gas.  We examine the relative effectiveness of acceleration by ultraviolet and infrared radiation fields, both independently and acting simultaneously on the same cloud.  We study clouds that are optically thin to infrared emission but with varying ultraviolet optical depths.  Consistent with previous work, we find relatively efficient acceleration and long cloud survival times when the infrared band flux dominates over the ultraviolet flux.  However, when ultraviolet is dominant or even a modest percentage ($\sim 5-10$\%) of the infrared irradiating flux, it can act to compress the cloud, first crushing it and then disrupting the outer layers.  This drives mixing of outer regions of the dusty gas with the hot diffuse background to the point where most dust is not likely to survive or stay coupled to the gas.  Hence, the cold cloud is unable to survive for a long enough timescale to experience significant acceleration before disruption even though efficient infrared cooling keeps the majority of the gas close to radiative equilibrium temperature ($T \lesssim 100$K).  We discuss implications for observed systems, concluding that radiation pressure driving is most effective when the light from star-forming regions is efficiently reprocessed into the infrared.

\end{abstract}


\section{Introduction}\label{sec:intro}

Galactic outflow are observed ubiquitously in star-forming galaxies \citep{2005ARA&A..43..769V}. In many cases, the observed outflow rates and velocities imply the outflows must have an important impact in the evolution of galaxies.  These outflows almost certainly play a role in regulating star-formation \citep[e.g.][]{2003ApJ...599...38B,2012MNRAS.422.2816B,2013MNRAS.428.2966P,2013Natur.499..450B}, but also affect the chemical evolution of galaxies as well as the circumgalactic and intergalactic medium \citep{2001ApJ...556L..11A,2008ApJ...674..151E,2008MNRAS.385.2181F,2010MNRAS.406.2325O}.

Multiphase winds have been observed in many star-forming galaxies, including molecular gas \citep[e.g.][]{2009ApJ...700L.149V,2014A&A...562A..21C,2017ApJ...835..265W,2017A&A...608A..38O,2018ApJ...864L...1G,2018Sci...361.1016S,2015ApJ...814...83L,2019arXiv190700731K}, neutral atomic gas \citep[e.g.][]{2000ApJS..129..493H,2002ApJ...570..588R,2005ApJ...621..227M,2016A&A...593A..30M,2018A&A...617A..38S}, and ionized gas \citep[e.g][]{1998ApJ...506..222M,1998ApJ...493..129S,2019MNRAS.487.3679M,2019MNRAS.488.1813T}.  Although not observed directly, constraints on hot gas outflows are provided by X-ray observations of star forming galaxies \citep{2007ApJ...658..258S,2014ApJ...784...93Z}.

It has generally been believed that these outflows are driven primarily by supernova feedback.  In high star formation rate galaxies, the overlapping supernova remnants merge to form giant bubbles of hot gas that break out of the cooler interstellar medium (ISM) gas.  The cooler gas is then entrained in these hot outflows \citep{1985Natur.317...44C,2009ApJ...697.2030S}.  However, it is not entirely clear that the much cooler gas will survive long enough to reach the inferred velocities due to shredding and mixing with the hotter background flow \citep{1994ApJ...420..213K,1990MNRAS.244P..26B,2015ApJ...805..158S,2018ApJ...854..110Z}. One possibility is that magnetic fields suppress the disruption due, to purely hydrodynamics instabilities \citep{2015MNRAS.449....2M,2016MNRAS.455.1309B} but it is unclear if the magnetic field strengths and geometries are present in such outflows.  Another possibility is that the gas is destroyed but condenses out of the hot flow due to radiative cooling at larger distances from the galaxy. \citep{2016MNRAS.455.1830T}.

It is possible that a number of different acceleration mechanisms play a role in launching outflows, with different mechanisms possibly dominating in different environments or a different stages in the acceleration \citep{2012MNRAS.421.3522H,2018Galax...6..114Z}.  In addition to entrainment, plausible mechanisms include radiation pressure of starlight on dust \citep[e.g.][]{2005ApJ...618..569M,2011ApJ...735...66M}, cosmic ray pressure \citep[e.g.][]{1975ApJ...196..107I,2008ApJ...687..202S,2017ApJ...834..208R,2019arXiv190301471W}, and active galactic nuclei \citep{2012ARA&A..50..455F,2014ARA&A..52..589H}.

In this work we focus on the role played by radiation pressure of starlight on dust. This mechanism has already been extensively studied with detailed radiation hydrodynamic numerical simulations at different scales and with varying assumptions and set-ups.  An important question has been the role of Rayleigh-Taylor instabilities in limiting the effectiveness of acceleration \citep{2012ApJ...760..155K,2013MNRAS.434.2329K,2015ApJ...809..187S,2016ApJ...829..130R}.  Despite the presences of such instabilities, it seems that some fraction of the radiation can be accelerated to large velocities \citep{2014ApJ...796..107D,2015MNRAS.453.1108T,2017ApJ...839...54Z} and may allow cold gas to survive longer than entrainment allows \citep{2018ApJ...854..110Z}. With the exception of \citet{2016ApJ...829..130R}, which studied the effect of radiation pressure in the local star cluster environment, most of these studies focus on infrared (IR) radiation pressure.  This is sensible for considering the role of radiation pressure on galactic scales since the vast majority of the light in the most extreme star-forming galaxies (luminous or ultraluminous infrared galaxies, hereafter LIRGs and ULIRGs) is reprocessed into the IR. However, most of the radiation originates from starlight radiated in the ultraviolet (UV) band.  It is possible that UV plays a greater role in the launching of gas close to the star clusters or in galaxies which lower dust obscuration. Indeed previous work has shown that UV radiation pressure may be important in the vicinity of massive star formation, although its role is limited by inhomogeneities in the gas arising from Rayleigh-Taylor instabilities \citep{2016MNRAS.463.2553R} and inhomegeneities in the radiation field itself from multiple stellar sources \citep{2018ApJ...859...68K}.  Although we neglect the impact of photoionization, but this may also be important for the destruction of molecular gas near starbursts \citep[e.g.][]{2016ApJ...819..137K}.

Therefore, we are motivated to consider the relative role played by UV and IR opacities in accelerating clouds.

The plan of this paper is as follows.  In section~\ref{sec:method} we describe our numerical simulation methods and problem set-up.  In section~\ref{sec:results} we report on the results of variety of simulations with differing assumptions about parameters of interest such as IR to UV flux ratios, optical depths as well as sensitivity to assumptions in the numerical method and simulation set-up.  We discuss the primary implications of our results in section~\ref{sec:discussion} and summarize our conclusions in section~\ref{sec:conclusion}.

\section{Method}\label{sec:method}
\subsection{Radiation Hydrodynamics Equations}\label{subsec:method_equation}

We solve the equations of hydrodynamic and radiation transfer using the {\sf Athena++} (Stone et al., in preparation) code.  The relevant equations are the (respectively) the equations of conservation of mass, momentum and energy:
\begin{eqnarray}\label{eq:coupledHD}
\frac{\partial\rho}{\partial t}+\nabla \cdot (\rho\mathbf{v})=0,\nonumber \\
\frac{\partial(\rho\mathbf{v})}{\partial t}+\nabla \cdot (\rho\mathbf{vv}+\mathsf{P})=-\mathbf{G},\nonumber\\
\frac{\partial E}{\partial t}+\nabla \cdot \left[(E+P)]\mathbf{v}\right]=-cG^{0}.
\end{eqnarray}
Here $\rho$, $\mathbf{v}$, $E$ and $P$ are fluid density, velocity, total energy density and pressure, $\mathsf{P}$ is the pressure tensor. The source terms $\mathbf{G}$ and $G_{0}$ represent the components of the radiation four force, which are calculated by taking moments of the radiation transfer (RT) equation. The total energy density is
\begin{eqnarray}
E=\frac{P}{\gamma-1}+\frac{1}{2} \rho v^2,
\end{eqnarray}
where $\gamma$ is the adiabatic index and the terms represent the gas internal energy and kinetic energy, respectively.  The radiation four-force is computed from the specific intensity $I_\nu$, which is evolved according to the time-dependent RT equation:
\begin{equation}\label{eq:RTtransfer}
    \frac{\partial I_{\nu}}{\partial t}+c\mathbf{n}\cdot\nabla I_{\nu} =  S_{\nu}(\mathbf{n}).
\end{equation}
Here $ S_{\nu}(\mathbf{n})$ is the radiation source term, $\mathbf{n}$ represents a unit vector parameterizing the direction, and $c$ is the speed of light. The RT equation is solved using an explicit-implicit scheme in Eulerian frame, similar to the method described in \citep{2014ApJS..213....7J}.  The main difference is that \cite{2014ApJS..213....7J} evaluate radiation source terms in the Eulerian frame by expanding to second order in $v/c$.  In the {\sf Athena++} implementations, the specific intensities are first transformed to the fluid comoving frame, where the opacities and emissivities are simplest.  The relevant source terms are evaluated and updated implicitly along with the comoving frame gas internal energy equation. The resulting source terms are integrated over frequency and angle and then transformed back to the Eulerian frame.

In this work, we integrate Equation~(\ref{eq:RTtransfer}) over frequency assuming the radiation field can be approximated with two radiation band representing infrared (IR) and optical/ultraviolet (UV) contributions to the radiation field. The resulting RT equations solved are:
\begin{eqnarray}
    \frac{1}{c}\frac{\partial \Iuv}{\partial t}+\mathbf{n}\cdot\nabla \Iuv & = & - \Gamma(\mathbf{n})\kapuv\rho \Iuv,\nonumber\\
    \frac{1}{c}\frac{\partial \Iir}{\partial t}+\mathbf{n}\cdot\nabla \Iir & = & \Gamma(\mathbf{n}) \kapir \rho\bigg(\frac{a_{r}T^{4}}{4\pi}-\Iir\bigg),
    \label{eq:transfer}
\end{eqnarray}
where $a_{r}$ is the radiation constant. The subscripts uv and ir label the ultraviolet and infrared opacities and radiation fields, respectively. Note that we have assumed there is no source of UV emission within the domain and that the UV radiation is only provide from an external source via the boundary conditions. The $\Gamma(\mathbf{n})$ accounts for transformations between the comoving and Eulerian frames and corresponds to
\begin{equation}
\Gamma(\mathbf{n}) = \gamma_{\rm L} \left(1 - \frac{\mathbf{v} \cdot \mathbf{n}}{c}\right),
\end{equation}
where $\gamma_{\rm L}$ is the Lorentz factor.  The specific intensities $\Iir$
and $\Iuv$ are evaluated in the Eulerian frame and the opacities $\kapir$ and $\kapuv$ are evaluated in the comoving frame.  However, we emphasize that the differences between the comoving and Eulerian frames are quite small in these simulations.  The radiation energy and momentum terms are computed by integrating the appropriate moments of Equation~(\ref{eq:transfer}) over angles.
\begin{eqnarray}
\label{eq:radmomemtumsource}
\mathbf{G} = \int d\Omega \mathbf{n} \Gamma(\mathbf{n}) \left[ \kapir \rho\frac{a_{r}T^{4}}{4\pi}- \kapuv\rho\left(\Iir+ \Iuv\right)\right],\\
  \label{eq:radenergysource}
G_{0} = \int d\Omega \Gamma(\mathbf{n}) \left[ \kapir \rho\frac{a_{r}T^{4}}{4\pi}- \kapuv\rho\left(\Iir+ \Iuv\right)\right].
\end{eqnarray}
In the limit of zero velocity, the corresponding momentum and energy source terms are the familiar expressions
\begin{eqnarray}
\mathbf{G} \rightarrow -\left(\frac{\kapuv\rho}{c}\mathbf{F}_{\rm uv}+\frac{\kapir\rho}{c}\mathbf{F}_{\rm ir}\right),\nonumber\\
G_{0} \rightarrow \kapir \rho (a_{r}T^{4}-E_{\rm ir})-\kappa_{\rm uv}\rho E_{\rm uv}.
\end{eqnarray}
Here $\mathbf{F}_{\rm ir}$ and $\mathbf{F}_{\rm uv}$ are the  IR and UV radiation flux, respectively, and $E_{\rm ir}$ and $E_{\rm uv}$ are IR and UV radiation energy density, respectively.

The UV dust opacity $\kapuv$ depends on dust grain sizes and species as well a frequency, but we use a constant representative value.   The IR opacity $\kapir$ is assumed to be a temperature dependent Rosseland mean opacity using the approximation of \citet{2012ApJ...760..155K}. To focus on the cold cloud dynamics, we ignore the scattering opacity, setting scattering opacity to zero. The dust opacity is
\begin{eqnarray}
\label{eq:dustopacity}
\kapir & = & 
    10^{-3/2}\left(\frac{T}{10\rm K}\right)^{2} s \; \rm{cm^{2}/g},\nonumber\\
\kapuv & = & 100 s \ \rm{cm^{2}/g}.
\end{eqnarray}
This assumes a Milky-Way-like dust-to-gas ratio and $\kapir$ is a reasonable approximation for $T\lesssim100$K and flattens at higher temperature\citep{2003A&A...410..611S}. Hence, we assume a constant value of $10^{1/2} \ \rm cm^2/g$ for $T > 100 \ \rm K$. The quantity $s$ represents a scaled ratio of dust-to-gas fraction, normalized so that $s=1$ corresponds to the initial dust to gas fraction (assumed to be uniform) in the cloud. The UV dust opacity will be dependent on both the grain size distribution and wavelength of the optical to UV spectral energy density of the source driving the cloud acceleration.  Our fiducial value is probably somewhat conservative in that dust opacity at UV wavelengths can be several times higher than this in the Milky Way ISM \citep[e.g][]{2011ApJ...732..100D}.  In our calculation, this may be offset by considering clouds with larger sizes or densities to make them more optically thick, although this choice does impact the ratio of IR to UV opacity, which may be an interesting parameter to explore in future work.

In order to track dust evolution, we initialize cold cloud gas with $s=1$ and cells in the hot background gas with $s=0$. We then evolve $s$ as a passive scalar via a continuity equation
\begin{equation}\label{eq:scalartransport}
    \frac{\partial s}{\partial t}+\mathbf{v}\cdot\mathbf{\nabla}s=0,
\end{equation}
which assumes that there is no source of dust other than the initial dust in the cloud. However, we adopt a simple prescription to account for the decoupling and destruction of the dust when it mixes with the hotter, less dense background gas, setting the passive scalar to zero for cells above a fiducial temperature of 1500K. This is approximately the temperature where most grain constituents are destroyed \citep{1994ApJ...421..615P}, resulting in drops in the opacity \citep{2003A&A...410..611S}. Mixed gas at this temperatures also typically has densities low enough that it is no longer clear that dust remains dynamically well-coupled with the gas, due to the increase in the mean-free-path of dust-gas collisions \citep{2013MNRAS.434.2329K}. A more sophisticated model of grain-gas interaction and grain destruction will be of interest in future studies but our simple scheme serves its primary purpose, which is to decouple hotter and more diffuse gas from the radiation field. We choose 1500K as a representative value but we have checked that decreasing our decoupling/destruction temperature to 500K or 1000K (where some dust constituents are destroyed) has no significant impact on our inferred survival times.

Since the transport portion of the transfer equation is solved explicitly, the Courant-Friedrichs-Lewy (CFL) condition in the code is set by the speed of light, which is much larger than the flow velocity or sound speed.  Hence, it is advantageous to adopt the reduced speed of light approximation, where $c$ in Equation~(\ref{eq:transfer}) is replaced by $\tilde{c} = R c$. Assuming $R \le 1$ allows one to take time steps that are a factor of $R^{-1}$ larger. As long as $R$ is not chosen to be too low, the time-dependent term remains small and the radiation flux close to quasi-steady on the flow timescale.

The conditions for validity of reduced speed of light approximation are described by \citet{2013ApJS..206...21S} in section 3.2. The main constraint is the need to preserve the correct ordering of characteristic timescales. The light-crossing (i.e. radiation diffusion) time should always be smaller than the dynamical time. The radiation travels at reduced light speed $\min(\tilde{c},\tilde{c}/\tau_{\rm max})$, where $\tau_{\rm max}$ is the maximum optical depth in the system. For a system with characteristic length $l_{0}$, $l_{0}/\min(\tilde{c},\tilde{c}/\tau_{\rm max})\ll l_{0}/v_{\rm max}$, where the dynamical timescale $l_{0}/v_{\rm max}$, $v_{max}$ is the velocity determines the dynamical timescale.  For the modest flow velocities and low optical depths considered here, these constraints are easily obeyed for $R = 0.01$.

\subsection{Simulation Setup}\label{subsec:method_initial}

We initialize all simulations with a cold dense cloud in pressure equilibrium with a hotter, less dense background gas. The cloud geometry is circular (2D) or spherical (3D) and it is initialized at rest in the center of the domain.  A summary of simulation parameters is provided in Table~\ref{tab:summary_params}. We initialize the cloud to fiducial temperature $ T_{0}=50$K.  We define a corresponding fiducial flux $F_{0}=c a_{r}T_{0}^{4}$. We first model the UV radiation from a galaxy or star-forming region within the galaxy as a constant uniform flux $F_{\rm uv}=1.4F_{0}\approx4.9\times10^{12}\rm L_{\odot}/kpc^{2}$, and inject the radiation flux from the bottom boundary.  We used an angular grid described by \citet{1999A&A...348..233B}. For 2D simulations, we used 6 angles per octans, yields 84 angles in total. For 3D simulation, we used 4 angles per octan, so it's 80 angles in total. The radiation flux in our simulation is scaled to the luminosity of ultraluminous infrared galaxies (ULIRGs). This choice of $F_{\rm uv}$ is about an order of magnitude smaller than typical IR radiation flux from ULIRGs \citep{2018ApJ...854..110Z}, which is higher than typically observed.  We choose a high value for our fiducial flux to provide favorable conditions for acceleration, but consider lower values in other calculations. In Section~\ref{sec:discussion} we discuss the impact of varying the radiation flux on cloud dynamics.

The incoming radiation field injected at the lower $x$ boundary is isotropic for incoming radiation on this boundary to model a distributed source of ultraviolet and infrared emission, as might be expected from a large starbursting region of a galaxy.  This is probably a good approximation for the infrared but the UV radiation may be more directed if the radiation field is dominated by a few relatively distant star clusters. The (half) isotropic radiation field allows optically thick clouds to be compressed in directions both parallel and perpendicular to the motion.  A more parallel directed radiation field might be expected compress the cloud primarily along the direction of motion \citep{2014ApJ...780...51P} leading to a more ``pancake'' structure in the initial evolution.

Given $T_{0}$, the fiducial speed $v_{0}$ is chosen to be the adiabatic sound speed $c_{\rm s}^2= k_{b} T_0/(\mu m_H)$, where $k_{b}$ is Boltzmann constant, and we assume the mean molecular weight $\mu=1.0$ for simplicity. \citet{2018ApJ...854..110Z} showed that varying $\mu$ has limited impact on cloud dynamics. For our fiducial run, we choose a fiducial initial density $\rho_{0}=1.0\times10^{-19}\rm g/cm^{3}$.  We applied a random perturbation $\delta\rho$ on the cloud density to make it moderately inhomogeneous, which $\delta\rho/\rho$ is randomly distributed between -0.25 and 0.25. We set the characteristic length scale $l_{0}$ to the initial cloud diameter $D_{\rm c}=0.411 \ \rm pc$, which corresponds to a column density of $N_{\rm H}=  7.566 \times 10^{22}\ \rm cm^{-2}$. The corresponding initial optical depths to IR and UV radiation are $\tau_{\rm ir}=0.100$ and  $\tau_{\rm uv}= 12.658$, so the cloud is optically thick to the UV radiation and optically thin to IR emission. The cloud is embedded in the interstellar medium with temperature $T_{\rm bkgd}=10^{5}T_{\rm c}=5\times10^{6}$K and the background density is lower by $10^{-5}$ to maintain pressure equilibrium.

With $T_{0}$ and $\rho_{0}$ defined, we introduce two dimensionless parameters
\begin{equation}
    \mathbb{P}=\frac{a_{r}T_{0}^{4}}{\rho_{0}v_{0}^{2}},\quad,\mathbb{C}=\frac{c}{v_{0}}
\end{equation}
$\mathbb{P}$ represents the ratio of radiation pressure and gas pressure, $\mathbb{C}$ represents the ratio of light speed and sound speed. With $\mathbb{P}$ and $\mathbb{C}$, the hydrodynamic equations and RT equation can be written in dimensionless form \citep{2014ApJS..213....7J}. The code solves the dimensionless versions of Equations~(\ref{eq:coupledHD}), (\ref{eq:RTtransfer}) and (\ref{eq:scalartransport}).

In the simulations, all the hydrodynamic boundaries are set to outflow boundary conditions.  Except for the lower $x$ boundary. All radiation boundary conditions are set to outflow, which copies the radiation fied in the last active zone, so that the radiation field in calculation domain is isotropic for rays aligned with the positive $x$-direction even close to the vertical boundaries. At the lower $x$ boundary, we impose a uniform and time-steady incoming radiation flux. 

In the 2D simulations, the vertical ($x$) and horizontal ($y$ or $z$) sizes of the simulation domain are $40D_{c}$ and $16D_{c}$ respectively, where $D_{c}$ is the radius of the cloud. However, the 3D simulation has smaller domain to reduce the computational cost. The size of simulation domain is $32D_{c}\times6.4D_{c}\times6.4D_{c}$, which spans $(-12D_{c},20D_{c})\times(-3.2D_{c},3.2D_{c})\times(-3.2D_{c},3.2D_{c})$ respectively. For all the simulations, in order to better resolve the cloud, we apply 3-level static mesh refinement to the central region $(-5D_{c},5D_{c})\times(-2D_{c},2D_{c})$ (in case of 3D simulation, the refined central region in $x$, $y$, and $z$ are $(-5D_{c},5D_{c})\times(-2D_{c},2D_{c})\times(-2D_{c},2D_{c})$ ). The resolution (cell size) at the most refined level is listed in Table~\ref{tab:summary_params}. We also enforce a pressure floor $P_{\rm floor}=10^{-5}\rho_{0}v_{0}^{2}$ and a density floor $\rho_{\rm floor}=10^{-5}\rho_{0}$. We use $R=10^{-2}$ and $C_{CFL}=0.4$ for all 2D runs except TLUV\_R, in which we use $R=10^{-3}$. In TLUV\_3D, we adopted $R=10^{-2}$ and $C_{CFL}=0.3$.

We find it useful to define the cloud mass as total mass of cold gas, which is tracked by dust and labeled by the passive scalar $s$. Then cloud mass is
\begin{equation}\label{eq:cloudmasswithin}
    M_{\rm{c}}\equiv\sum_{i}s_{i}\rho_{i}V_{i}
\end{equation}
where $i$ runs over every grid cell in the simulation domain and $V_i$ is the volume of cell $i$.

Initially (at $t=0$), the passive scalar is set to be $s=1$ within the cloud, and $s=0$ in the background material. Hence, $M_{\rm c}$ is a representation of the  mass of dusty gas within the calculation domain. At later times, the cloud mass is the initial mass $M_{\rm c,0}$ minus the accumulative overheated gas mass:
\begin{equation}
M_{\rm c}(t) = M_{\rm c,0}-\int \dot{M}_{\rm loss}(t)dt
\label{eq:cloudmasssubtract}
\end{equation}
where $\dot{M}_{\rm loss}(t)$ represents the sum of all gas that has exited the domain or been lost to mixing with the hotter background gas.  Note that the latter mechanism (mixing with hot gas) is the dominant loss channel in all simulations.

In order to focus on cloud evolution, we adopted a cloud-following frame approach, so the center of mass of the cloud remains fixed in the calculation domain.

In the cloud following scheme, the $x$ component of the mean velocity of dusty gas is computed at the end of every time step as
\begin{equation}\label{eq:vmean}
  \Delta v_{\rm mean}=\frac{\int v_{x} \rho s dV}{\int \rho s dV}.
\end{equation}
Then $\Delta v_{\rm mean}$ is subtracted from $v_x$ for every cell in the simulation domain.  These boosts are then summed to keep track of the velocity $v_{\rm mean}$ of the total cloud velocity after each time step.  The hydrodynamics of the cloud is unaffected by these boosts due to the Gallilean invariance of the underlying hydrodynamic equations. In contrast, the radiation equations are not Galilean invariant but Lorentz invariant.  Hence, the radiation intensities differ at second order in $v/c$ from the true Eulerian frame.  For the calculations presented here, these discrepancies remain quite small and have almost negligible impact on our results.

\subsection{A Simple Model}\label{subsec:cloudmodel}

A characteristic hydrodynamical timescale is set by the sound crossing time:
\begin{equation}
    t_{0}=\frac{D_{c}}{v_{0}}\approx 6.255\times10^{5}\left(\frac{50 \rm K}{T_0}\right)\left(\frac{D_{c}}{0.411 \rm pc}\right) \ \rm{yr}.
\end{equation}
There are several radiation timescales of interest.  The first is the bulk acceleration timescale.  We estimate this by ignoring the detailed geometry and assuming the cloud is an opaque rectangle with opacity $\kapuv$, density $\rho_0$ ,length $L_x \sim l_0$, and uniform UV flux $F_{\rm uv}$ along $x$ direction. The radiation attenuates in the opaque cloud as $F_{UV}e^{-\kapuv \rho_0 x}$. The equation of motion is then
\begin{equation}\label{eq:cloudmodelEoM}
    \int \rho_0 \frac{d v}{dt} \  dx \ dA = \int \frac{\kapuv \rho_0}{c} F_{\rm uv}e^{-\kapuv \rho_0 x} \ dx \ dA
\end{equation}
Integrating Equations~(\ref{eq:cloudmodelEoM}) and assuming that the cloud is rigidly accelerated gives the average acceleration
\begin{equation}\label{eq:acc}
a = \frac{d\langle v \rangle}{dt}=\frac{\kapuv F_{\rm uv}}{c}\frac{1-e^{-\tau_{\rm uv}}}{\tau_{\rm uv}},
\end{equation}
with $\tau_{\rm uv}=\kapuv \rho_0 l_0$. When $\tau_{\rm uv} > 1$, Equation~(\ref{eq:acc}) yields a characteristic acceleration rate $a\approx \kapuv F_{\rm uv}/(c \tau_{\rm uv}) \equiv a_{\rm uv}$.

The UV radiation field heats the cold cloud while pushing on it.  This heating is predominantly balanced by cooling via the IR radiation, so we estimate the approximate cloud characteristic temperature $T_{\rm eq}$ by setting the UV absorption rate to balance the IR emission rate in the optically thin cloud:
\begin{eqnarray}
  a_{r}\kapir(T_{\rm eq})T_{\rm eq}^{4} & = & \kapuv E_{uv}+ \kapir(T_{\rm eq}) E_{ir},\nonumber\\
  & = & 2\left(\frac{\kapuv F_{uv}}{c}+\frac{\kapir(T_{\rm eq})F_{ir}}{c}\right).
  \label{eq:tequil}
\end{eqnarray}
The factor of 2 at RHS comes from moment integration of a half isotropic (isotropic for rays with $\mathbf{n} \cdot \hat{x} > 0$ and zero for rays with $\mathbf{n} \cdot \hat{x} < 0$) radiation field to get energy density and flux in $x$ direction. The resulting value is $T_{\rm eq} = 153.8$K for the fiducial run.

The radiation field doesn't just accelerate and heat the cloud, but also acts to compress it.  This is particularly true when the dusty cloud is opaque to the UV radiation and there is significant radiation pressure gradient across the cloud.  If we neglect the internal pressure support of the cloud, the time to crush it is simply determined by the relative acceleration of the cloud surface relative to the cloud center.  With this assumption, we define a radiation crushing timescale as
\begin{equation}\label{eq:radcrushingtime}
t_{\rm rad}=\sqrt{\frac{D_{c}}{\Delta a_{\rm uv}}}
\end{equation}
where $\Delta a_{\rm uv}=\kapuv F_{\rm uv}(1-e^{-\tau_{\rm uv}/2})/c$ represents the radiation acceleration difference between the outer radius of the the cloud and its center due to the self-shielding of the UV flux.

\section{Results}\label{sec:results}

We preformed a series of 2D and 3D simulations to study various factors that impact on the cloud dynamics. We list the relevant parameters used in all simulations in Table~\ref{tab:summary_params}. First we report the fiducial run TLUV in Section~\ref{subsec:result_fiducial}, which is an opaque cloud accelerated by pure UV radiation flux. Parameters in this run were chosen with the expectation that they would provide favorable conditions for cloud survival.  Next, we describe the impact of varying the optical depth in Section~\ref{subsec:result_uvthin}. Since the cloud responds differently to IR and UV radiation fluxes, we also report on the impact of varying the ratio of IR to UV flux in Section~\ref{subsec:result_irflux}. We discuss the dependence of ours results on dimensionality, resolution, and our choice for the reduced speed of light in Section~\ref{subsec:result_3D}.

\begin{deluxetable*}{lccccccc}
\tablecolumns{8}
\caption{Summary of Simulation Parameter}
\label{tab:summary_params}
\tablehead{
\colhead{Name} & \colhead{$\tau_{IR}$} & \colhead{$\tau_{UV}$} & \colhead{$F_{UV}$\tablenotemark{a}}  & \colhead{$F_{IR}$\tablenotemark{a}} & \colhead{$T_{\rm eq}$\tablenotemark{b}} & \colhead{$D_{c}$ (pc)} &  \colhead{Resolution (pc)\tablenotemark{c}}
}
\startdata
TLUV & 0.1 & 12.658 & $2.0\times10^{3}$ & 0.0 & 3.075 & 0.411 &  $[4.110\times10^{-3}]^{2}$\\
TSUV\_L & 0.01 & 1.266 & $2.0\times10^{3}$ & 0.0 & 3.075 & 0.041 &  $[4.110\times10^{-4}]^{2}$\\
TSUV\_D & 0.01 & 1.266 & $2.0\times10^{3}$ & 0.0 & 3.075 & 0.411 &  $[4.110\times10^{-3}]^{2}$\\
TSUV\_DL & 0.001 & 0.127 & $2.0\times10^{3}$ & 0.0 & 3.075 & 0.041 &  $[4.110\times10^{-4}]^{2}$\\
TLIR\_E & 0.1 & 12.658 & 0.0 & $2.0\times10^{3}$ & 1.297 & 0.411 & $[4.110\times10^{-3}]^{2}$\\
TLIR\_H & 0.1 & 12.658 & 0.0 & $2.0\times10^{4}$ & 2.306 & 0.411 & $[4.110\times10^{-3}]^{2}$\\
TLMF\_10 & 0.1 & 12.658 & $2.0\times10^{3}$ & $2.0\times10^{4}$ & 3.293 & 0.411 & $[4.110\times10^{-3}]^{2}$\\
TLMF\_5 & 0.1 & 12.658 & $1.0\times10^{3}$ & $2.0\times10^{4}$ & 2.922 & 0.411 & $[4.110\times10^{-3}]^{2}$\\
TLMF\_1 & 0.1 & 12.658 & $2.0\times10^{2}$ & $2.0\times10^{4}$ & 2.469 & 0.411 & $[4.110\times10^{-3}]^{2}$\\
TLUV\_3D & 0.1 & 12.658 & $2.0\times10^{3}$ & 0.0 & 3.075 & 0.411 &  $[8.220\times10^{-3}]^{3}$\\
TLUV\_HR & 0.1 & 12.658 & $2.0\times10^{3}$ & 0.0 & 3.075 & 0.411 & $[2.055\times10^{-3}]^{2}$\\ 
TLUV\_LR & 0.1 & 12.658 & $2.0\times10^{3}$ & 0.0 & 3.075 & 0.411 &  $[8.220\times10^{-3}]^{2}$\\
TLUV\_R & 0.1 & 12.658 & $2.0\times10^{3}$ & 0.0 & 3.075 & 0.411 &  $[4.110\times10^{-3}]^{2}$\\
\enddata

\tablenotetext{a}{Flux in units $\rm erg/s/cm^{2}$}
\tablenotetext{b}{$T_{\rm eq}$ in units of $T_{0}=50$K}
\tablenotetext{c}{The number in bracket gives the resolution at the finest level for each run in pc, the superscript outside bracket gives dimension of corresponding simulation.}

\end{deluxetable*}

\subsection{UV Optically Thick Cloud}\label{subsec:result_fiducial}

\begin{figure*}
    \centering
    \includegraphics[width=0.9\textwidth]{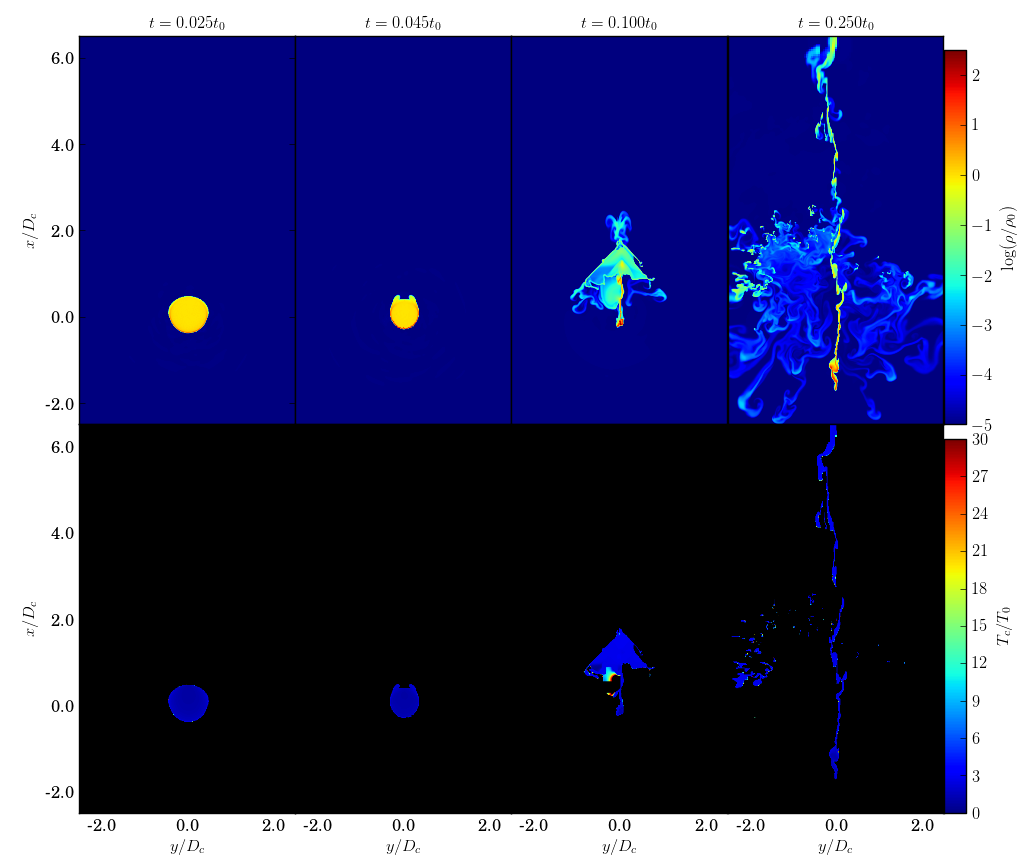}
    \caption{Simulation snapshots from TLUV. Top panels: density snapshots of both cold and hot gas. Lower panels: temperature of cold dusty gas, the background medium is masked by black. The maximum temperature in color bar corresponds to the temperature at which we set the the passive scalar to zero, representing the overheating of cold gas. $t_{0}=6.255\times10^{5}$yr, $D_{c}=0.411$pc, $\rho_{0}=10^{-19}\rm g/cm^{3}$ and $T_{0} =50$K.}
    \label{fig:result_fidsnap}
\end{figure*}

The TLUV run examines an optically thick cloud irradiated by a large UV radiation flux. This setup assumes a rather large UV flux that would only be possible relatively close to a very compact, high star formation rate region within a luminous galaxy.  The column and gas density are also quite large, so much so that it would be unstable to gravitational collapse if self-gravity were modeled here.  A large column is chosen because we would like to explore the optical depth effects on the cloud.  This combination might be expected to be favorable to cloud acceleration (large UV flux) and survival (significant self-shielding).

In this run, $\tau_{\rm ir}=0.1$, $\tau_{\rm uv}=12.7$. The incoming radiation flux from the lower $x$ boundary is pure UV flux. Figure~\ref{fig:result_fidsnap} shows density and temperature snapshots from this simulation. The first row shows density of both hot and cold gas, the second row shows only temperature of dusty gas $(s \ne 0$). Before radiation flux intact with the cloud, the cloud cools rapidly to IR radiative equilibrium, leading to an abrupt initial pressure drop. However, well before the cloud responds to the pressure mismatch with background medium, the incoming radiation field sweeps through the cloud. Immediately after radiation reaches the cloud, the cloud temperature rises to around the estimated equilibrium temperature $T_{\rm eq}$.

At early times $(t < 0.1t_{0})$, the opaque cloud is compressed by the radiation pressure gradient within the cloud, which causes the side of the cloud facing the radiation field to be accelerated more strongly than the side opposite this face. Since the cloud is optically thick to UV radiation, radiation only directly acts on the gas near the cloud surface. A dense distorted front is formed, and Rayleigh-Taylor-like instabilities grow at the interface between the hot and cold gas. The interior of the cloud is shielded from radiation and stays cold. As the radiation continues compressing the cloud, the gas pressure increases, with the cloud reaching its volume minimum near $t\sim0.1t_{0}$.

After this point the gas pressure gradient counterbalances radiation pressure, the cloud partially re-expands and loses its initial spherical symmetry. In the re-expansion phase, the dense core of the cloud remains cold ($T \sim T_{\rm eq}$) and is stretched slightly, primarily along the direction of motion. The lower density envelope of gas becomes turbulent and filamentary.  Although the center of mass of the dusty gas remains fixed on the grid, the boosts associated with the cloud following scheme result in a significant velocity in the background medium towards the bottom $x$ boundary, roughly at the value of $v_{\rm mean}$ in Figure~\ref{fig:uvthin_vvdisp}.  The combination of the large relative velocity of the background flow and the radiation pressure from the UV drive Kelvin-Helmholz like instabilities that facilitate the mixing with the hotter background gas. On any single time step, a small fraction of the gas is heated above the assumed destruction temperature. The detailed morphology of low density gas at late times is sensitive to the assumed initial condition, which we tested by considering different random perturbations, but the qualitative picture of compression, re-expansion, and mixing outlined here was qualitatively similar in all runs.

In Figure~\ref{fig:uvthin_vvdisp}, the black solid line shows the evolution of the cloud mean velocity $v_{\rm{mean}}$ (top panel), velocity dispersion $\sigma_{v}$ (middle panel) and cold gas mass $M_{\rm c}$ (bottom panel). The black dashed lines in the first row is the velocity evolution corresponds to a constant acceleration at $a$ (Equation~[\ref{eq:acc}]). Despite the simplicity of the model, $a$ provides a good estimation of cloud bulk acceleration within one radiation crushing time. At around $t\approx0.1 t_{0}$, the cloud acceleration drops relative to this prediction.  The cloud enters the re-expansion phase and the velocity dispersion $\sigma_{v}$ increases.  The lower density outer layers of the cloud begin to mix with the background, exceeding 1500K and the dust is assumed to be destroyed ($s$ is set to zero). Hence, the cloud mass begins to drop significantly even though outflow through the simulation boundary remains low.  We stop the simulation at $t=0.4t_{0}$, when outflow through the boundaries starts to become significant, but a large fraction of the initial cold cloud mass has already mixed with the background.

The black solid line in Figure~\ref{fig:uvthin_avg} is the evolution of mean density $\bar{\rho}$ and temperature $\bar{T}$ of cold gas in TLUV. These represent mass and dust weighted averages via $\bar{X} \equiv \int X \rho s dV/\int \rho s dV$. The average density rises as the radiation keeps compressing the cloud, with most of the mass in the dense core, which is strongly compressed. This mass weighted average density persists at a level higher than $\rho_{\rm init}$ even as the cloud begins to partially reexpand because most of the mass remains in the dense core.  Note that this mass weighted average emphasizes the density in the cloud core, but the volume average density actually drops as the large extended envelope of low density gas expands. The average temperature of the cold gas is slightly below the estimated equilibrium temperature $T_{\rm eq}$ in the compression phase, and gradually rises toward $\bar{T} \approx T_{\rm eq}$ in the re-expansion phase, but never quite reaches it.  This is primarily a problem with our $T_{\rm eq}$ estimate, which assumes the energy density and flux our related by $E_{\rm uv} = 2 F_{\rm uv}/c$.  The energy density is actually lower than this due primarily to the attenuation of the incoming UV and (minor) differences in the assumed angular distribution, leading to a slightly smaller $T_{\rm eq}$ than our estimate predicts.  The rise in temperature can be attributed primarily to more gas being exposed to the incident radiation field as the cloud re-expands but mixing with the hotter background plays a role as well.

The average temperature is always well below the dust destruction temperature, indicating that only a small fraction of the dusty gas in the envelope is mixing with the background on any time step. This suggests there is a continuous flux of cooler, higher density gas leaving the compressed cloud core for the envelope, mixing with the background, and being destroyed.  As discussed in section~\ref{subsec:destruction} this process happens via mixing on the radiation hydrodynamic timescales.  Efficient IR cooling prevents the radiation from simply over heating the cloud on the radiation crossing time.

\subsection{UV Optically Thin Clouds}\label{subsec:result_uvthin}

As noted above, the TLUV run corresponds to a rather large cloud mass and column.  Here we consider UV optically thin runs, with lower cloud columns that might be more typical of outflowing gas in star forming environments.   We studied two runs with optical depth of about unity, both with $\tau_{\rm uv}=1.266$, a factor of 10 lower than TLUV.  The UV optical depth $\tau_{\rm uv}=\kapuv \rho_{0}D_{c}$, can be made smaller by either reducing the radius or lowering the density. The cloud with lower density is in TSUV\_D, the cloud with smaller radius is in TSUV\_L. For a third run, TSUV\_DL, we reduce both the density and length, giving $\tau_{\rm uv}=0.127$. In all cases, the optical depth to IR radiation $\tau_{IR}$ is reduced by the same factors.
\begin{figure}
    \centering
    \includegraphics[width=0.5\textwidth]{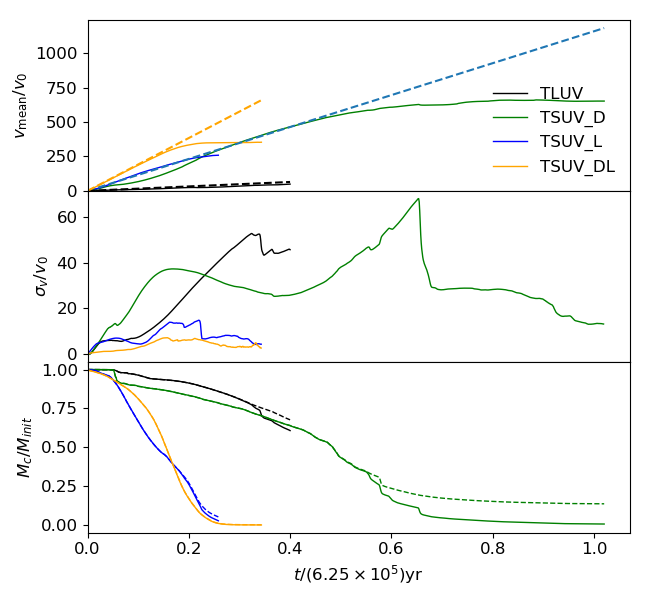}
    \caption{Mean velocity $\Delta v_{\rm mean}$ (top panel), velocity dispersion $\sigma_{v}$ (middle panel) and cloud mass $M_{\rm{c}}$ (bottom panel) evolution of TLUV (black), TSUV\_D (green), TSUV\_L (blue) and TSUV\_DL(orange). In the top panel, the dashed line with the same color is the time integration of $a$ (Equation~\ref{eq:acc}) of each run, excepting TLUV\_D and TLUV\_L has the same $a$. In the bottom panel, solid lines are $M_{\rm c}$, dashed line with the same color is $M_{\rm c}(t)$ of each run. We ended the simulations when $M_{\rm c}$ and $M_{\rm c}(t)$ starts to diverge, meaning that cold gas exiting the simulation box starts to effect total mass loss. $v_{0}=0.642$km/s.}
    \label{fig:uvthin_vvdisp}
\end{figure}

Figure~\ref{fig:uvthin_vvdisp} compares the cloud bulk motion of optically thin clouds to the optically thick cloud in TLUV. In both TSUV\_L (blue) and TSUV\_D (green), estimated acceleration in Equation~(\ref{eq:acc}) (shown as a dashed curve) is about 10 times larger than TLUV.  Both optically thin runs roughly follow this curve, falling slightly below during the re-expansion stages. The compression and re-expansion is qualitatively similar to what is seen in the TLUV run, leading to similar evolution of velocity dispersion and cloud mass, albeit on somewhat different timescales.  For example, cloud mass is already substantially depleted by $t \simeq 10^5$ years in the TSUV\_L and TSUV\_DL runs. The TSUV\_DL run is accelerated by greater amount consistent with the expectations from Equation~(\ref{eq:acc}) and the velocity dispersion tends to be lower but the mass loss initially follows the TSUV\_L run and is destroyed on a comparable timescale.  The mass loss in the TSUV\_D run follows similar evolution to the TLUV run, but we run it for longer because it takes longer for outflow through the domain boundaries to dominate dust mass loss.  We see that the runs with shorter lengthscale are disrupted much faster, consistent with the shorter cloud crushing timescale implied by equation (\ref{eq:radcrushingtime}). Note that the drops in velocity dispersion near the end of runs follows in part from no longer being able to track mass leaving the domain.

\begin{figure}
    \centering
    \includegraphics[width=0.5\textwidth]{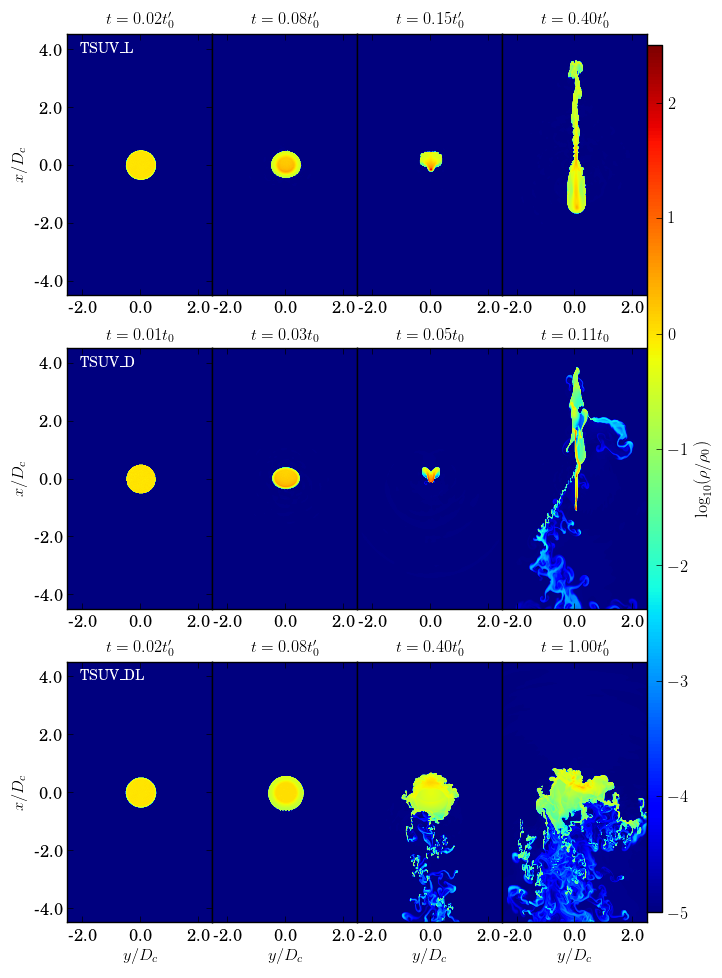}
    \caption{Density snapshots of TSUV\_L (top panels) ,TSUV\_D (middle panels) and TSUV\_DL (bottom panels), the cloud is more optically thin compared to TLUV in these runs. The first row: TSUV\_L, where the cloud has smaller diameter $D_{c}=0.1l_{0}$. The second row panel: TSUV\_D is the cloud with lower density $\rho=0.1\rho_{0}$. The third row: TSUV\_DL, the cloud has both lower density and smaller radius. Notice that the $D_{c}$ of TSUV\_L and TSUV\_DL are different than TSUV\_D. $t_{0}=10t_{0}'=6.255\times10^{5}$yr}
    \label{fig:result_uvthin}
\end{figure}

Figure~\ref{fig:result_uvthin} compares the density snapshots of the optically thin clouds. The morphology and evolution of TSUV\_L and TSUV\_D are qualitatively similar. Due to the smaller optical depth, the radiation is more uniformly distributed than in the TLUV run. The clouds do not show a dense front at cloud-radiation interface like in TLUV, where the radiation is absorbed. Hence, the clouds are more uniformly compressed by radiation pressure at early times. As the clouds re-expand, the core of the dense gas is somewhat stretched along the direction of motion.  As in TLUV, lower density gas pushed by both radiation pressure and the interaction with the background gas leading to a turbulent, filamentary structure.   The low density regions ejected by the cloud mix with hot background, gradually heating the gas to the destruction temperature.  The TSUV\_DL run follows a similar evolution at early times, but there is much less overall compression because the radiation field is much more uniform.  In fact, the average density in the cloud drops as the outer layers expand.  Nevertheless, the subsequent evolution is qualitatively similar to the re-expansion phases of the other runs.  Shear at the interface with the background ISM disrupts the cloud and drives mixing with the hotter background and heats the cold gas to $T >1500$ K.

\begin{figure}
    \centering
    \includegraphics[width=0.5\textwidth]{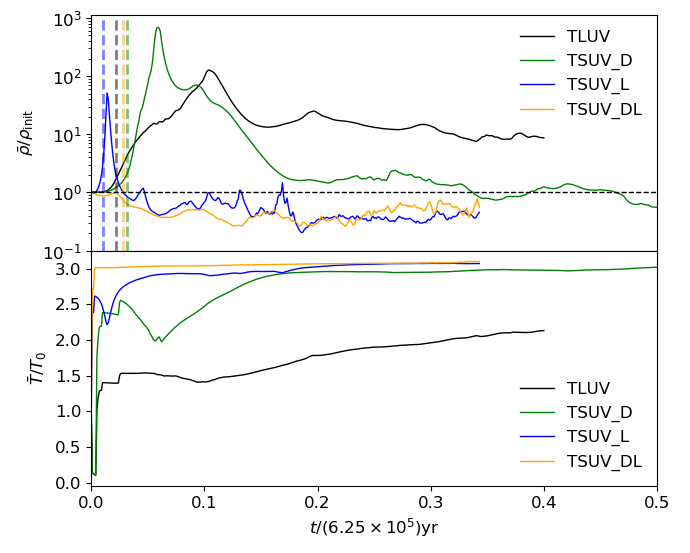}
    \caption{Average density (top panel) and temperature (bottom panel) weighted by cold gas density for TLUV (black), TSUV\_D (green), TSUV\_L (blue) and TSUV\_DL (orange). $\rho_{\rm init}=\rho_{0}=10^{-19}\rm g/cm^{3}$. TSUV\_D and TSUV\_DL have lower $\rho_{\rm init}=0.1\rho_{0}=10^{-20}\rm g/cm^{3}$. $T_{0}=50$K. The vertical dashed lines in the first row is the radiation crushing time $t_{\rm rad}$ for corresponding simulations. Notice that the X axis is scaled to $6.255\times10^{5}$ yr.}
    \label{fig:uvthin_avg}
\end{figure}

Figure~\ref{fig:uvthin_avg} also shows the difference in their average density $\bar{\rho}$ and temperature $\bar{T}$ for the optically thin runs. As already noted, the TSUV\_L run is compressed faster than the runs with larger length scales, but has lower maximum density.  The TSUV\_D run is also compressed slightly faster than the TLUV run, but reaches a much higher maximum average density. Note that TSUV\_D has both lower cloud and background density, so the density and temperature contrast between the cloud and background is the same as TLUV. As noted above the TSUV\_DL run does not show the compression and re-expansion behavior that is seen in the more optically thick runs. The average density drops continuously. During the initial compression phase, the TSUV\_L and TSUV\_D runs remain below $T_{\rm eq}$ estimate due to a lower $E_{\rm uv}$ in the cloud core, but rise to near this value at later times as the optically thin assumption underlying this estimate holds better in these runs than in TLUV.
The evolution is roughly similar to the optically thin simulations in \citet{2014ApJ...780...51P}, but with larger relative velocity between the cloud and background.

\begin{figure}
    \centering
    \includegraphics[width=0.5\textwidth]{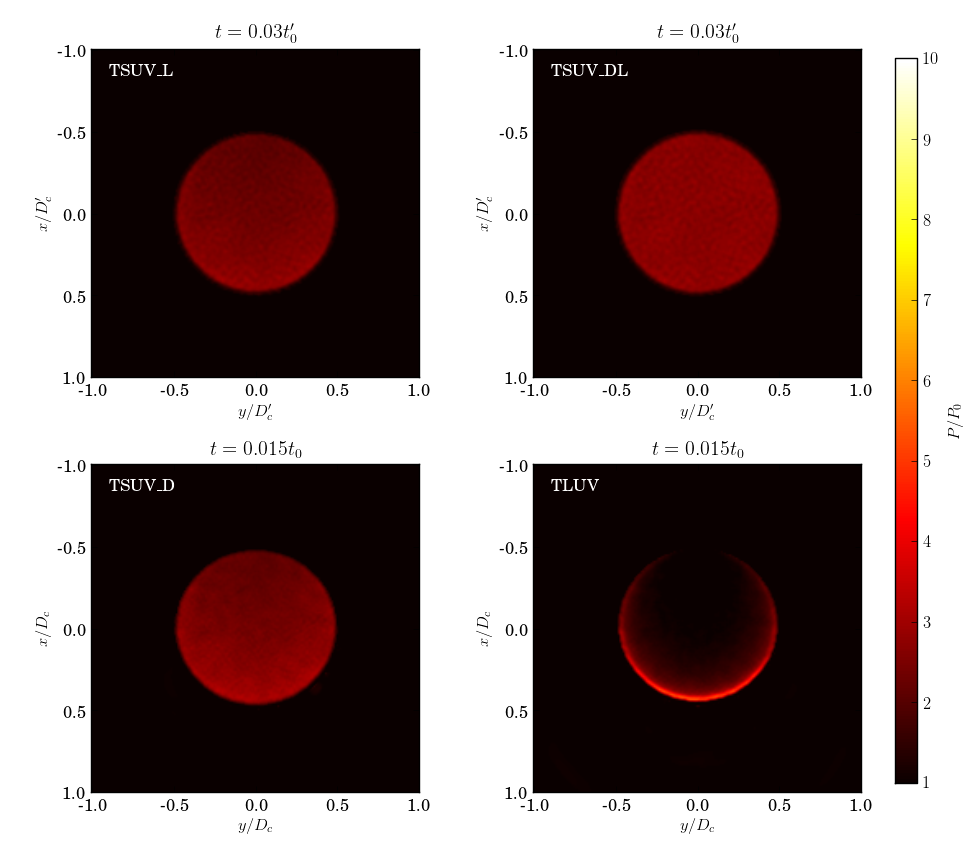}
    \caption{Gas pressure distribution of TSUV\_L (upper left), TSUV\_DL (upper right), TSUV\_D (lower left) and TLUV (lower right) at same compression stage. $t_{0}'=0.1t_{0}$, $D_{c}'=0.1D_{c}$. Characteristic pressure $P_{0}=\rho_{0}v_{0}^{2}$, $P_{0}=4.127\times10^{-10}\rm dyne/cm^{2}$ for TSUV\_L and TLUV, $P_{0}=4.127\times10^{-11}\rm dyne/cm^{2}$ for TSUV\_D and TSUV\_DL.}
    \label{fig:uvthin_pressure}
\end{figure}

The differences in evolution between optically thick and optical depth unity runs can be attributed primarily to the gas pressure distribution within the clouds. Initially, both clouds are in pressure equilibrium with the background medium. After radiation has swept through the cloud, different parts of cloud experience different radiation forces.  The radiation pressure gradients are modest in the optically thin runs but self-shielding leads to strong gradients in the UV optically thick run.  These radiation pressure gradients compress the cloud until a comparable gas pressure gradient develops to oppose it. Figure~\ref{fig:uvthin_pressure} shows the pressure snapshots of TSUV\_L, TSUV\_D, TSUV\_DL  and TLUV at the same compression stage. TSUV\_L and TSUV\_D are at optical depths of unity, so the gas pressure gradient is relatively small.  In contrast, the gas pressure is strongly enhanced near the surface in the optically thick TLUV runs (right).  The outward pressure gradient forces supports the clouds and slow down the compression.  TSUV\_DL is even more optically thin and show almost no pressure gradient at early times.

Initially, these gas pressure effects are modest and our estimate of the cloud crushing time $t_{\rm rad}$ in Equation~(\ref{eq:radcrushingtime}) yields a good order of magnitude estimate the time for both the optically thick and optical depth unity clouds to reach their maximum average density.  However, if we look more quantitatively we can see the impact of the gas pressure gradient force, which is not accounted for Equation~(\ref{eq:radcrushingtime}). Our estimates of $t_{\rm rad}$ are shown as vertical dashed lines in Figure~\ref{fig:uvthin_avg}. They provide better estimates of the time when the cloud reaches peak density in the two optical depth unity runs, but the maximum compression of the optically thick cloud is slightly delayed due to the resistance from the gas pressure gradient. Note that compression is somewhat faster in the TSUV\_L run because $t_{\rm rad}\propto\sqrt{D_{c}}$ and $D_{c}$ is smaller in this run. Modest compression does occur in the TSUV\_DL run, but it doesn't show up in this average density plot because the outer envelope of the cloud expands by a greater amount than the core of the cloud contracts.

Since the temperature is similar in both optical depth unity runs, the lower initial density in the TSUV\_D run means that it has a lower initial gas pressure than the TSUV\_L run. Since the cloud reaches its peak density when the gas pressure gradient becomes large enough to support the cloud against radiative compression, it has largest peak average density $\bar{\rho}_{\rm peak}/\rho_{\rm init}$ (green solid line) in Figure~\ref{fig:uvthin_avg}.

\subsection{Acceleration with Both IR and UV Irradiation}\label{subsec:result_irflux}

In contrast to UV radiation flux, cloud acceleration with IR radiation is generally gentler due to the smaller optical depth and the fact that IR radiation acts both to compress the cloud (incident radiation) but also provides a support against compression (re-radiated IR). Previous work has suggested that clouds accelerated solely by IR radiation might survive longer than cloud entrained in a hot outflow \citep{2018ApJ...854..110Z}.  Hence, we have performed a number of runs with an incident IR flux to compare with the pure UV results discussed above.  We first consider two purely IR runs: TLIR\_E with an incident IR flux equal to the UV runs above and TLIR\_H, which has a flux a factor of 10 larger. For both cases, we set the initial IR optical depth  $\tau_{\rm ir}=0.1$.

\begin{figure}
    \centering
    \includegraphics[width=0.5\textwidth]{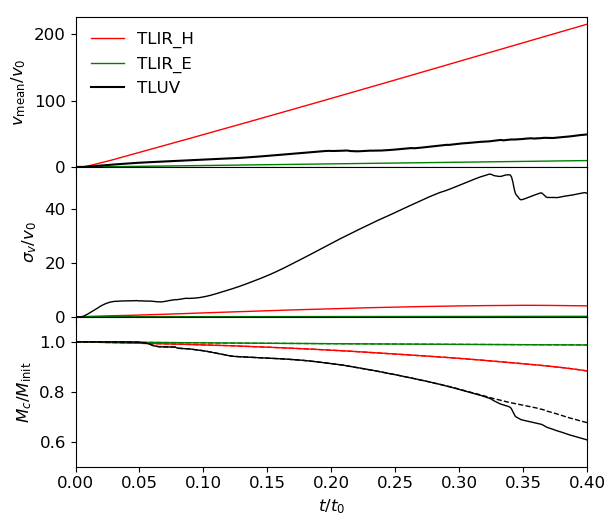}
    \caption{Mean velocity $\Delta v_{\rm mean}$ (top panel), velocity dispersion $\sigma_{v}$ (middle panel) and cloud mass $M_{\rm{c}}$ (bottom panel) evolution of of TLIR\_H (red), TLIR\_E (green) and TLUV (black). $v_{0}=0.642$km/s, and $t_{0}=6.255\times10^{5}$yr}
    \label{fig:ir_vvdispmass}
\end{figure}

Figure~\ref{fig:ir_vvdispmass} shows the cloud mean velocity $v_{\rm{mean}}$ (the first row), velocity dispersion $\sigma_{v}$ (the second row) and cold gas mass $M_{\rm c}$ (the third row) for TLIR\_E (green) and TLIR\_H (red), with TLUV (black) included for comparison. TLIR\_E is accelerated much more slowly than TLUV due to the smaller opacity, which gives rise to an acceleration $a_{\rm ir} \approx \kapir F_{\rm ir}/c < a_{\rm uv}$ when $F_{\rm ir} = F_{\rm uv}$.  Since the IR driven cloud is not compressed significantly, the cloud is more weakly disrupted and retains its initial structure longer, leading to lower velocity dispersion.  The mean acceleration is nearly constant, giving rise to a nearly linear velocity profile.  A similar evolution is seen for TLIR\_H, but the factor of 10 increase in $F_{\rm ir}$ compensates for the lower opacity and ultimately leads to a more rapid acceleration than seen for TLUV.

\begin{figure}
    \centering
    \includegraphics[width=0.5\textwidth]{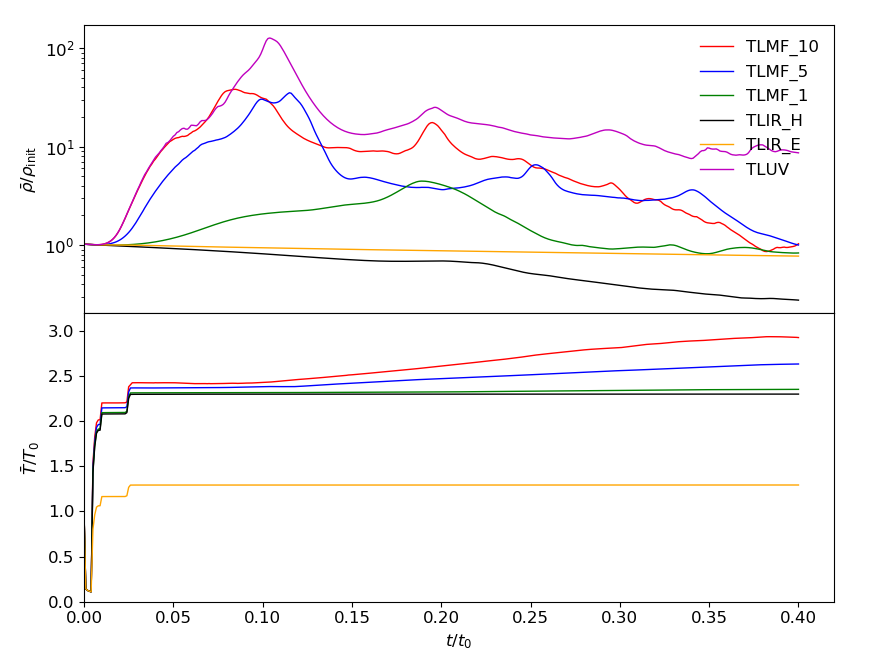}
    \caption{Average density (top panel) and temperature (bottom panel) of cold gas for multi-frequency runs TLMF\_10 (red), TLMF\_5 (blue), TLMF\_1 (green) and infrared radiation runs TLIR\_H (black), TLIR\_E (orange).  $\rho_{\rm init}=\rho_{0}=10^{-19}\rm g/cm^{3}$, $T_{0}=50$K.}
    \label{fig:2freq_avg}
\end{figure}

Figure~\ref{fig:2freq_avg} shows the cold gas average density (upper panel) and temperature (lower panel) for TLIR\_H (black) and TLIR\_E (orange). In contrast to the UV runs, both the mean density and temperature remain relatively constant for these runs.  In fact, the average density shows a slight drop as the radiation pressure associated with the re-emitted IR leads to the cloud becoming weakly over-pressured relative to the background and expanding slightly. There is a brief initial transient when the radiation sweeps across the cloud and heats it to the equilibrium temperature, where it remains for the rest of the evolution.  Overall, our results are in good agreement with the large scale optically thin cloud simulations performed by \citet{2018ApJ...854..110Z}.

Given the disparate evolution histories and survival times in the IR-only and UV-only runs presented thus far, it is natural to ask how a combination of UV and IR driving affects the cloud evolution. For highly star-forming galaxies, such as ULIRGs, the UV usually represents a small fraction $\lesssim 1$\% of the total observed emission. Most of this radiation is thought to be originally emitted in the optical and UV by stars, and then reprocessed in the IR due to the large dust optical depths along most lines of site.  However, it is plausible the UV will have been less attenuated in the location where the outflows are launched, motivating an exploration of different ratios of UV to IR flux.

\begin{figure}
    \centering
    \includegraphics[width=0.5\textwidth]{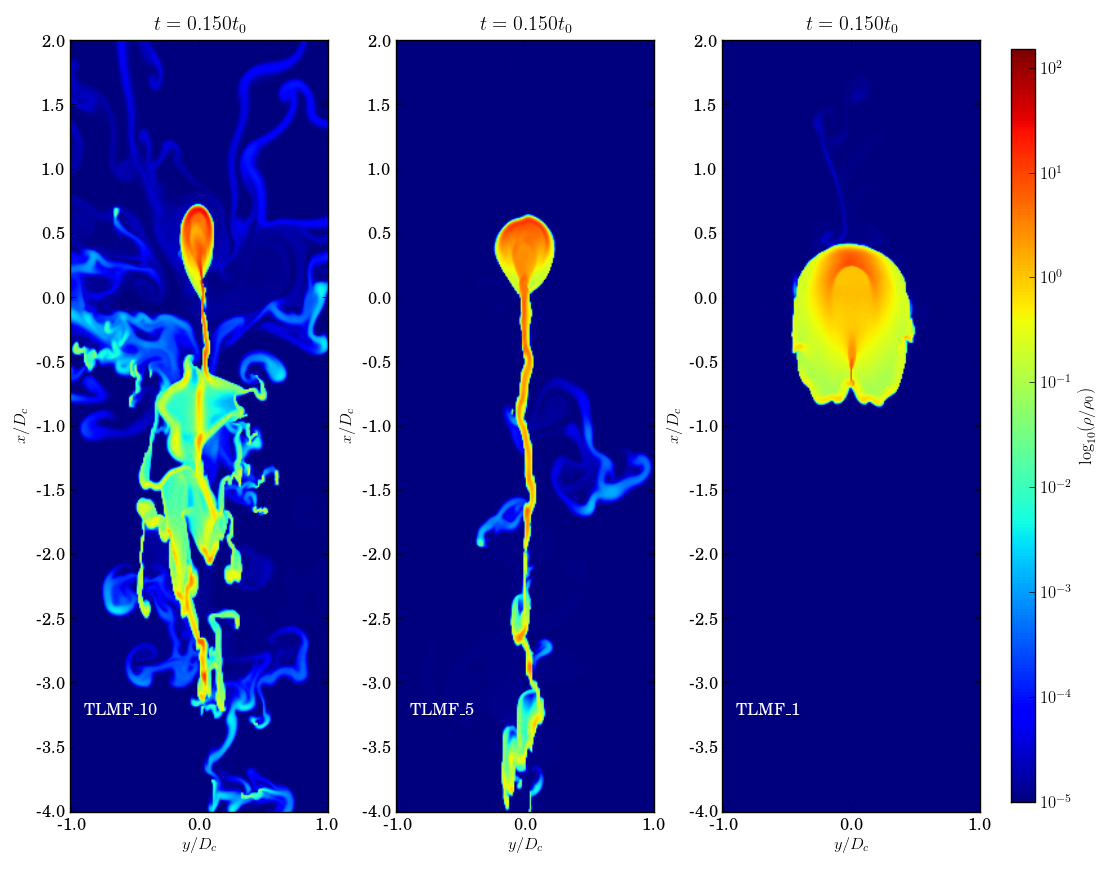}
    \caption{Density snapshots of TLMF\_10 (left panel), TLMF\_5 (middle panel) and TLMF\_1(right panel) at $t=0.15t_{0}$. Here $t_{0}=6.255\times10^{5}$yr, $D_{c}=0.411$pc, $\rho_{0}=10^{-19}\rm g/cm^{3}$}
    \label{fig:TLMFdensity}
\end{figure}

We consider three simulations all with the same incident IR flux, which is equivalent to TLIR\_H. These simulations also have an incident UV flux corresponding to $1\%$ (TLMF\_1) , $5\%$ (TLMF\_5), and $10\%$ (TLMF\_10) of the IR flux. The parameters are listed in Table~\ref{tab:summary_params}. The cloud is optically thick to UV radiation ($\tau_{\rm }=12.7$) and optical thin to IR radiation ($\tau_{IR}=0.1$).

\begin{figure}
    \centering
    \includegraphics[width=0.5\textwidth]{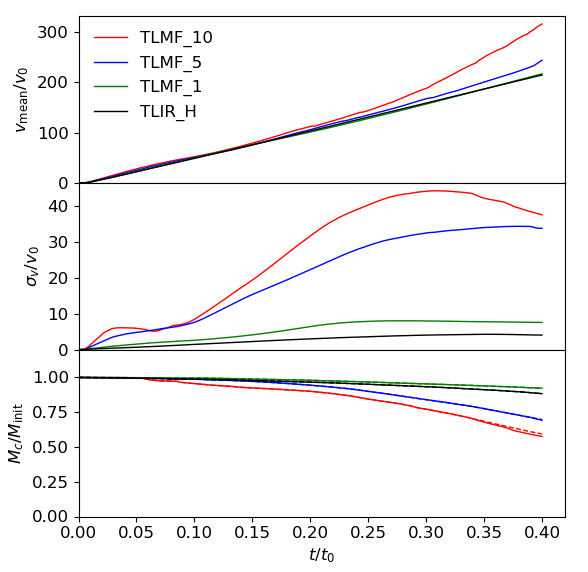}
    \caption{Mean velocity $\Delta v_{\rm mean}$ (top panel), velocity dispersion $\sigma_{v}$ (middle panel) and cloud mass $M_{\rm{c}}$ (bottom panel) evolution of TLMF\_10 (red), TLMF\_5 (blue), TLMF\_1 (green), TLIR\_H (black) . $v_{0}=0.642$km/s, and $t_{0}=6.255\times10^{5}$yr}
    \label{fig:2freq_vvdispmass}
\end{figure}

As with UV only runs, the UV provides a differential radiation force that acts to first compress the cloud until gas pressure rises and drives re-expansion. This density evolution is seen most clearly for TLMF\_5 and TLMF\_10 in Figure~\ref{fig:2freq_avg}, but is only modest for TLMF\_1, which is similar to the IR only runs.  Density snapshots shortly after maximum compression (at $0.15 t_0$) are shown in Figure~\ref{fig:TLMFdensity}, where one can see the increasing degree disruption as the UV fraction increases. It's also interesting to compare the average cold gas density of TLMF\_10 and TLUV in Figure~\ref{fig:2freq_avg}. Despite the same incident UV flux, TLMF\_10 has lower peak density than TLUV because of the support from re-emitted IR radiation against UV radiation. As the cloud is heated, average temperature increases, TLMF\_1 approaches the optical-thin equilibrium temperature $T_{eq}$ estimate, but both TLMF\_5 and TLMF\_10 lie below (Table~\ref{tab:summary_params}). As in the TLUV run, the UV radiation energy density inside the cloud is lower than our estimate would imply due to attenuation within the UV optically thick cloud.

Figure~\ref{fig:2freq_vvdispmass} shows the mean velocity (the first row), velocity dispersion (the second row), and mass evolution (the third row) for these with the pure IR run (TLIR\_H) for comparison. As we add more UV flux, there is a slight enhancement in the acceleration at early times but the effect is mostly modest for the lower two runs, with the strongest enhancement coming at later time in the run with 10\% UV flux after cloud has already been substantially disrupted. This evolution is also responsible for the increasing velocity dispersion associated with the cloud disruption as the UV fraction increases.  As in the UV only runs, the mixing leads to substantial losses of dusty gas in the TLMF\_5 and TLMF\_10 runs, but the TLMF\_1 run is similar to the IR-only runs.  The evolution of the TLMF\_10 run is qualitatively similar to the TLUV run, which has the same incident UV flux, and about half the dusty cold gas is overheated by $t=0.4 t_0$, when outflow becomes significant.

The overall impression is that the acceleration is enhanced if there is enough UV flux so that the radiation force from UV matches or exceeds the IR radiation force.  However, the survival of dusty gas is sensitive to the relative contributions of UV and IR radiation that drives it. Since the temperature of the majority of the gas remains close to the equilibrium temperature (see Table~\ref{tab:summary_params} and equation~[\ref{eq:tequil}]), it is not a matter of the UV directly heating the gas, but instead driving dynamical evolution of the cloud that enhances mixing with the hotter background gas. For low UV radiation fraction case like TLMF\_1, the presence of IR might helps the support the cloud against the differential acceleration from UV radiation, limiting the effects of the UV. At higher UV radiation fraction in cases like TLMF\_10 and TLMF\_5, the dynamical effect from UV radiation dominates the cloud evolution by triggering compression and re-expansion, leading to mixing and cloud destruction similar to the pure UV runs.

\subsection{Dimensionality, Resolution, and Reduced Speed of Light}\label{subsec:result_3D}

\begin{figure}
\centering
\includegraphics[width=0.5\textwidth]{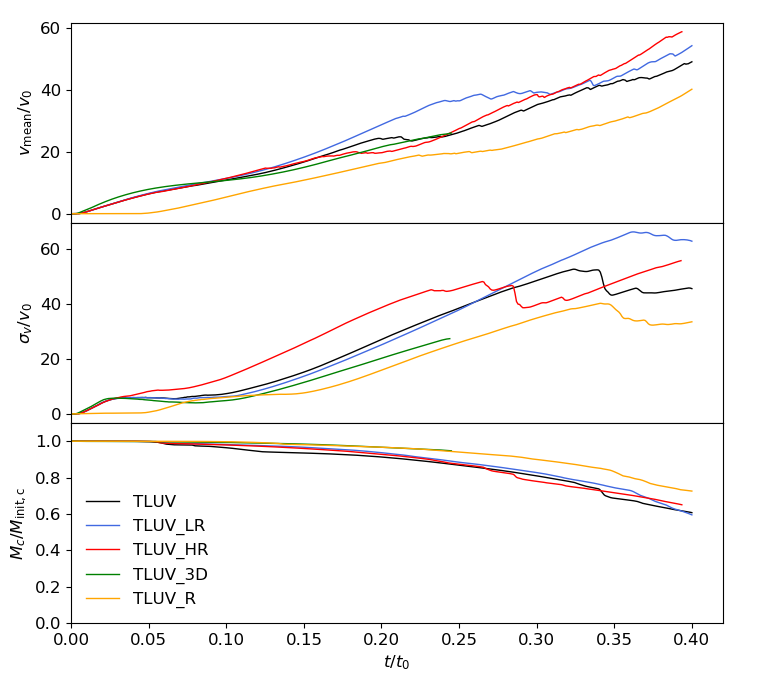}

\caption{Mean velocity $\Delta v_{\rm mean}$ (top panel), velocity dispersion $\sigma_{v}$ (middle panel) and cloud mass $M_{\rm{c}}$ (bottom panel) evolution of TLUV\_3D (green), TLUV (black) ,TLUV\_LR (blue), TLUV\_HR (red), TLUV\_R (orange). In TLUV\_R, the radiation flux travels 10 times slower than other runs because of lower reduction factor. $v_{0}\approx0.64$km/s, $t_{0}\approx6.255\times10^{5}$yr.}
\label{fig:res_vvidspmass}
\end{figure}

\begin{figure}
    \centering
    \includegraphics[width=0.5\textwidth]{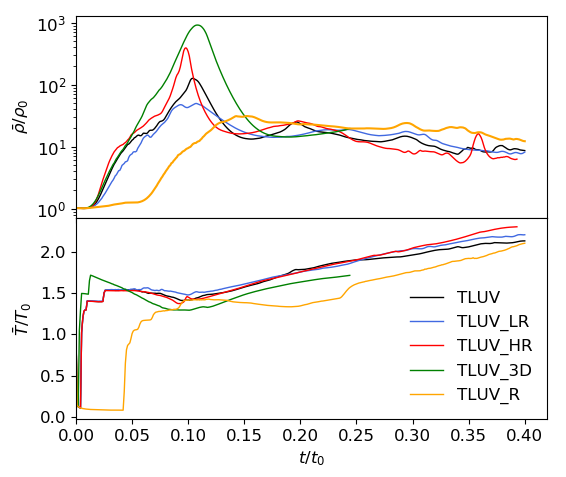}
    \caption{Average density (upper panel) and temperature (lower panel) of cold gas in TLUV\_3D (green), TLUV (black) ,TLUV\_LR (blue), TLUV\_HR (red), TLUV\_R (orange). $\rho_{\rm init}=\rho_{0}=10^{-19}\rm g/cm^{3}$, $T_{0}=50$K. In TLUV\_R, the radiation flux travels 10 times slower than other runs because of smaller reduction factor, we moved the curves of TLUV\_R $0.01t_{0}$ earlier in order to compare the cloud dynamics with other runs.}
    \label{fig:result_res_avg}
\end{figure}

The hydrodynamic interactions that drive mixing and cloud destruction are potentially sensitive to resolution.  We considered the effects of resolution by rerunning our fiducial run at two additional resolutions.  The TLUV\_HR and TLUV\_LR runs are performed at resolution that is a factor of two higher and lower, respectively, in both dimensions relative to the TLUV run (see Table~\ref{tab:summary_params}). Figure~\ref{fig:res_vvidspmass} shows the bulk motion and mass evolution of the cloud for different resolutions. Evolution of the mean density and temperature are shown in Figure~\ref{fig:result_res_avg}.

The motion and mass evolution of the runs are all qualitatively similar to each other, although there are modest deviations in later evolution, when the non-linear effects start to dominate. These later time deviations are at about the same level as we see when changing the random initial perturbations on the cloud density.  Slightly more sensitivity is seen in the evolution of the average density, where the maximum average density reached scales with resolution, suggesting that the core cloud structure is not yet resolved at peak.  However, almost all runs asymptote to similar values of density at late times and follow show little variation in temperature evolution. Hence, our results do not seem to be substantially impacted by resolution for the conditions considered here.

All the simulations presented above were 2D, but the hydrodynamic effects that lead to mixing with the background might depend on dimensionality so we also performed a 3D simulation.  Since 3D runs are considerably more expensive we only carry out one run (TLUV\_3D) to see how well our 2D results generalize to 3D.  Due to the increased computational cost, the resolution of TLUV\_3D is chosen to be equivalent to the TLUV\_LR run. Comparing these runs in Figure~\ref{fig:res_vvidspmass}, we find that the acceleration and survival times are rather similar for both runs.

\begin{figure}
    \centering
    \includegraphics[width=0.5\textwidth]{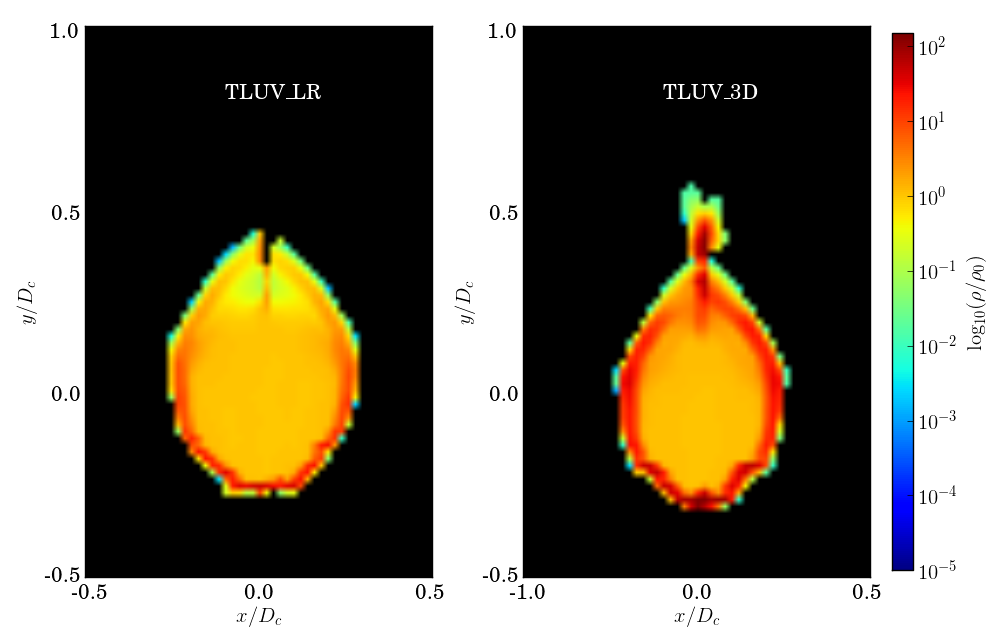}
    \caption{Dust density snapshots of TLUV\_LR (Left) and TLUV\_3D (Right)  at $t=0.06t_{0}$, the hot background medium is masked by black. $t_{0}=6.255\times10^{5}$yr, $\rho_{0}=10^{-19}\rm g/cm^{3}$.}
    \label{fig:result_3ddens}
\end{figure}

Figure~\ref{fig:result_3ddens} compares a 2D slice from TLUV\_3D with a density snapshot at the same time in the TLUV\_LR run.  The compression of the 3D cloud is qualitatively consistent with the compression of the 2D cloud, with slightly higher compression occurring near the surface where the UV flux is absorbed.  Comparison of mean density in Figure~\ref{fig:result_res_avg} indicates that the 3D run experiences somewhat higher maximum compression, which is consistent with the cloud being compressed nearly homologously in three rather than two dimensions since our half isotropic radiation field However, as with resolution, the simulations asymptote to similar densities at later times and we conclude that dimensionality has relatively little effect on the cloud acceleration or survival time for this setup.

Finally, we also test the effect of the speed of light reduction factor $R$ on cloud dynamics. In TLUV\_R, we chose a reduction factor of $R=10^{-3}$. In other words, the speed of light is 10 times smaller than the TLUV run, allowing for time steps that are 10 times larger. As a result it takes 10 times longer for the radiation from the lower $x$ boundary to reach the cloud. Once this offset is accounted for, the cloud bulk motion, velocity dispersion, mass (Figure~\ref{fig:res_vvidspmass}), and mean density and temperature evolution (Figure~\ref{fig:result_res_avg}) are all qualitatively similar to TLUV. The primary difference is that the peak average density is lower in this run, which is to be expected since the sound speed is closer to the reduced speed of light, this allows for more gradual adjustments in the gas pressure as radiation sweeps across the cloud.   Nevertheless, the density asymptotes to similar mean cloud densities at later times.  Hence, we do not believe our simulations are sensitive to our choice for $R$.

\section{Discussion}\label{sec:discussion}
\subsection{Destruction Mechanism for Cold Gas}
\label{subsec:destruction}

Our primary interest in this problem is assessing whether radiative acceleration could play an important role in accelerating the outflows observed in molecular and atomic transitions.  Therefore, an important constraint is that the gas cannot be too hot for the observed transitions to be present.  Since the optimal temperature ranges for different species can vary significantly, there is no single temperature cutoff that describes all transitions. We have utilized the presence of dusty gas $s > 0$ as the criterion for survival. Since we have a adopted 1500K as our dust destruction/decoupling temperature this can be thought of as a proxy for molecular gas.  We do not believe our results are significantly sensitive to this choice of temperature because the bulk of the cloud stays close to the radiative equilibrium temperature $T \lesssim 100$K, and only a modest fraction is a temperatures significantly higher than this. The dusty gas (defined by $s > 0$) at temperature significantly higher than the equilibrium temperature is being rapidly heated by mixing with the background gas to temperatures near the assumed background temperature ($T \gtrsim 10^5$K). For this reason test runs with a lower destruction/decoupling temperature (500K or 1000K instead of 1500K) are not significantly different because the gas reaching 500K continues heating and quickly exceeds 1500K shortly after.  Hence, our results should not be sensitive to the assumed destruction temperature as long as this temperature is well below the background temperature and well above the equilibrium temperature.

As discussed in Section~\ref{sec:results}, the process of cloud disruption and heating is primarily a radiation hydrodynamical rather than simply radiation transfer process.  In other words, the cloud is not simply heated to high temperature by the ultraviolet flux.  The efficient infrared dust cooling allows the majority of gas to remain close to the radiative equilibrium temperature.  Instead, the radiation pressure forces drive compression and re-expansion of the gas in optically thick clouds.  In the optically thin limit, the compression and re-expansion phase is much more subdued, but the dynamics still leads to mixing of the outer layers. In all simulations, a significant fraction of lower density dusty gas in the outer envelope of the cloud is heated by this mixing with the hotter background ISM resulting in destruction of most of the cloud on relatively short timescales.

\begin{figure}
    \centering
    \includegraphics[width=0.5\textwidth]{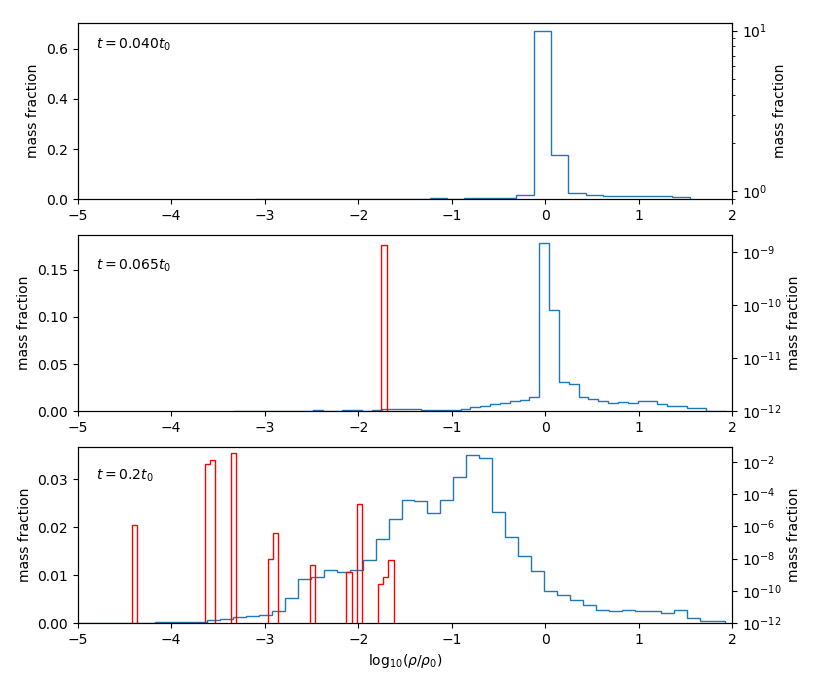}
    \caption{Cold gas density distribution of TLUV at $t=0.04t_{0}, 0.075t_{0}, 0.11t_{0}$ (the first, second, third row respectively). The distribution of all dusty material in the calculation domain is the blue solid line (with labels at left). The distribution of hot material, which we defined as material that with temperature higher than $95\%$ of the assumed dust destruction temperature, is the red solid lines (with labels at right).}
    \label{fig:discuss_densavg}
\end{figure}

Figure~\ref{fig:discuss_densavg} shows the density distribution of hot and cold gas in the cloud ($s\neq 0$) for the TLUV run at three different times. The blue solid lines are the density distribution of cold gas (left y-axis labels). The red solid lines are the density distribution of hot gas (right y-axis labels), which we defined as gases with temperature higher than $95\%$ of dust destruction temperature. The first row is before compression, the absence of red solid line means at $t=0.04t_{0}$, there is no hot dusty gas. The second row is before the volume minimum, where the mean cloud density is near its extremum. The hot dusty gas (red) has a density about 2 order of magnitude smaller than majority of cold dusty gas (blue). The bottom panel shows the density distribution during the re-expansion, with hot gas still corresponding to lower densities than most of the cold gas. In effect, there is a continuous flux of cold dense gas towards lower densities due to expansion and mixing that is heated and incorporated into the background ISM.

After the cloud begins to be disrupted, the implied rapid mixing with the background ISM is consistent with expectations from purely hydrodynamic models, driven by Kelvin-Helmholtz instabilities \citep{1990MNRAS.244P..26B}. We assume a mixing time scale $t_{\rm mix}\sim t_{\rm KH}\sim(\rho_{c}/\rho_{h})^{1/2}t_{\rm edd}$, with $\rho_{c}$ the density of cold material, and $\rho_{h}$ the density of hot material, and $t_{\rm edd}$ is the timescale for eddies to cross mixing layer. We estimate $t_{\rm edd}\sim 0.1l_{0}/\sigma_{v}$ with $\sigma_{v} \sim 10 - 50$ across different runs.  For the relatively low density envelope of dusty gas, we have ratios  $\rho_{c}/\rho_{h}\sim 10^{2}-10^{3}$, yielding $t_{\rm mix} \approx 0.02t_{0} - 0.3 t_{0}$.  Therefore it is not surprising that gas driven to lower densities by hydrodynamic processes in the cloud surface can rapidly mix with the background. However, if we apply the same argument to the initial cloud with density ratio $10^5$ times larger than the background, the timescale is much longer, indicating that radiative forces play an important role in the disruption.

\subsection{Cloud Survival Time}\label{subsec:diss_surviving}

A number of numerical studies of purely hydrodynamic entrainment \citet{2015ApJ...805..158S,2016ApJ...822...31B,2018ApJ...854..110Z} concluded that entrainment in a hot wind is unlikely to accelerate the cloud to the observed speed before the cloud is shredded.  However, \citet{2018ApJ...854..110Z} found that clouds accelerated by an IR radiation field can survive longer if a sufficiently larger infrared flux is available to accelerate them.  Our focus here was to consider the degree to which the addition or substitution of UV radiation impacts this conclusion. 

\begin{figure}
    \centering
    \includegraphics[width=0.5\textwidth]{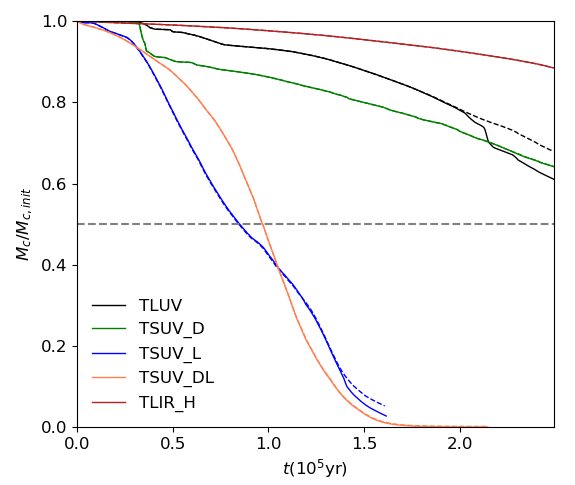}

    \caption{Cloud mass evolution for different runs. The horizontal grey dashed line labels when cloud mass is half of initial mass, corresponding to the cloud surviving time. Black lines are for TLUV. The blue lines are TSUV\_L, green lines are TSUV\_D, and orange lines are TSUV\_DL. Red lines shows mass evolution of the optical thick cloud in TLIR\_H, which is accelerated by pure IR radiation. For each color, the solid line is $M_{\rm c}$ (Equation~\ref{eq:cloudmasswithin}), the dashed line is corresponding $M_{\rm c}(t)$ (Equation~\ref{eq:cloudmasssubtract}).}
    \label{fig:discuss_survivaltime}
\end{figure}

Following \citet{2018ApJ...854..110Z}, we define the cloud surviving time as the time when the cloud lost half of its initial mass. Our definition of cloud mass follows from Equation~(\ref{eq:cloudmasswithin}), which includes mass loss both from mixing and advection out of the domain. In Figure~\ref{fig:discuss_survivaltime}, the solid lines correspond to $M_{\rm c}$, the dashed line with same color excludes mass loss associated with advection through the domain boundary. In other words it assumes (conservatively) that cold gas advected through the boundary remains cold and only gas overheated within the domain is accounted for.  Since we stop the simulations when the mass loss out of the domain begins to become significant, almost all of the mass loss shown in the figures is due to mixing and overheating within the simulation domain.

A potentially important characteristic timescale for estimating the cloud survival time is the radiation crushing time $t_{\rm rad}$ from Equation~(\ref{eq:radcrushingtime}).  For runs with $\tau_{\rm uv} \ge 1$ the mass loss occurs primarily after maximum compression, when the cloud starts to re-expand and the timescale for re-expansion is comparable to or slightly longer than $t_{\rm rad}$. Hence, for this regime it provides an approximate estimate of the survival time but the dependence on $\tau_{\rm uv}$ is not borne out. It implies that the most optically thick clouds will crushed the fastest but this is not what we found.  The difference arises because $t_{\rm rad}$ neglects the impact of the pressure support from gas and reradiated IR radiation.

Since the UV radiation acts to compress the cloud without providing pressure support, one might expect the optically thick cloud accelerated by UV radiation alone to have shorter surviving time than one supported by IR alone and this is consistent with our results. Figure~\ref{fig:discuss_survivaltime} compares the cloud mass evolution of the UV driven runs that are  optically thick (TLUV) to optically thin (TSUV\_D ,TSUV\_L and TSUV\_DL) with the IR driven run (TLIR\_H). The cloud mass drops significantly faster when accelerated by UV radiation. For clouds with optical depths of order unity or smaller, one expects the higher opacity to UV radiation to result in a  more rapid acceleration (see equation~[\ref{eq:acc}]). However, the significantly shorter survival time still limits the velocity to less than is typically inferred from observations.

With the exception of the pure IR runs (which are run for the same time as the fiducial run), we end all the simulations when outflow out of the simulation domain starts to become the dominant mass loss mechanism.  At this point, TLUV only has a speed $v_{\rm mean}\approx31.5$km/s, while $M_{\rm c}/M_{\rm c,init} \approx 60\%$, much smaller than the observed velocities, which are 100s of km/s \citep{2017A&A...608A..38O,2019arXiv190700731K}. Integrating the mean velocity over time we can estimate the ``flying'' distance $z\sim\int v_{\rm mean}dt \approx 3.7$pc, which is small compared to the typical size of star forming region. Therefore, we do not expect optically thick UV-driven clouds to survive long enough to explain observed outflows. TSUV\_L has larger acceleration and therefore reaches larger velocities of $\sim 131.4$ km/s by $t_{\rm final}$, but flying distance remains small due to the short survival time. The most promising UV run is TSUV\_D, which survives longer than TSUV\_L and reaches velocities of $\sim 368.0$ km/s when the mass drops to $\sim 36\%$ of its initial mass.

\begin{deluxetable}{lccc}
\tablecolumns{4}
\caption{Flying distance and final velocity}
\label{tab:distancevel}
\tablehead{
\colhead{Name} & \colhead{z (pc)} & \colhead{$v_{\rm final}$ (km/s)} & \colhead{$t_{\rm final}$ (yr)\tablenotemark{a}}
}
\startdata
TLUV & 3.691 & 31.462 & $2.502\times10^{5}$ \\
TSUV\_D & 61.269 & 367.992 & $3.314\times10^{5}$\\     
TSUV\_L & 7.348 & 131.369 & $1.099\times10^{5}$\\
TSUV\_DL & 10.839 & 196.716 & $1.071\times10^{5}$\\
TLIR\_E & 0.748 & 6.190 & $2.502\times10^{5}$\\
TLIR\_H & 16.647 & 138.029 & $2.502\times10^{5}$\\
TLMF\_10 & 20.147 & 203.361 & $2.502\times10^{5}$\\
TLMF\_5 & 17.542 & 156.901 & $2.502\times10^{5}$\\
TLMF\_1 & 16.560 & 139.504 & $2.502\times10^{5}$\\
\enddata
\tablenotetext{a}{For TSUV\_D, TSUV\_L and TSUV\_DL, which evolve long enough so that the cloud mass drops below $10\%$ of its initial mass, $t_{\rm final}$ is defined as the time when the cloud mass drop to $e^{-1.0}$ of initial mass. For the rest of simulations, $t_{\rm final}$ is the time we stop the simulation. $z$ and $v_{\rm final}$ are the flying distance and mean velocity corresponding to $t_{\rm final}$ respectively. }
\end{deluxetable}

In contrast, IR radiation flux accelerates clouds that are optically thin to IR for several dynamical timescales with relatively little disruption \citep{2018ApJ...854..110Z}. Incident radiation momentum is converted efficiently into cloud momentuum.  Although there is still some turbulent motion in the cloud outer layers and associated mixing with the background, it is much less than in the UV and the cloud survival time is much longer.

\subsection{Effects of Multiband Irradiation}

\begin{figure}
    \centering
    \includegraphics[width=0.5\textwidth]{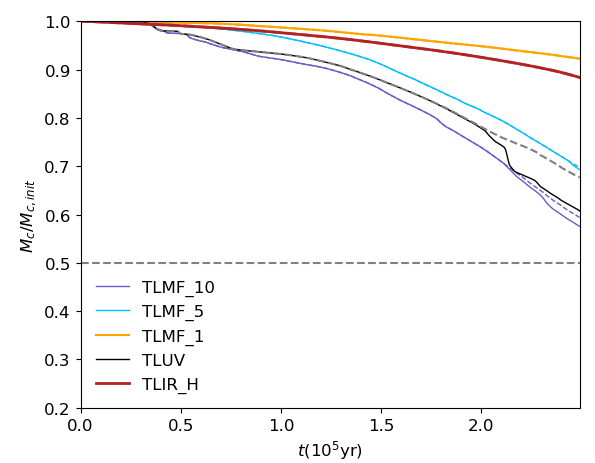}
    \caption{Cloud mass evolution for multi-frequency runs. The horizontal grey dashed line labels half of cloud initial mass, so the time reach it corresponds to the cloud surviving time. The red lines are TLIR\_H, the black lines are TLUV. The purple lines are the multi-frequency flux TLMF\_10, its UV flux is the same as TLUV and IR flux is the same as TLIR\_H. Then we fix the IR flux value, lower the UV flux fraction to $5\%$ of IR flux in TLIR\_5 (light blue), $1\%$ of IR flux in TLIR\_1 (orange). The solid lines are Equation~(\ref{eq:cloudmasswithin}) and the dashed lines are Equation~(\ref{eq:cloudmasssubtract}).}
    \label{fig:discussion_mf_surviving_time}
\end{figure}

Since a cloud absorbs both UV and IR radiation, but only re-emits IR radiation (Equation~\ref{eq:radenergysource}), the UV and IR radiation interact very differently with the clouds. UV radiation tends to accelerate the opaque cloud faster but also compresses the cloud and eventually drives greater mixing with the hot background ISM.  In contrast, the IR radiation flux accelerates the cloud more uniformly without significant compression, leading mixing to occur on much longer timescales. A cloud's interaction with a mixture of IR and UV flux is somewhat more complicated. Figure~\ref{fig:discussion_mf_surviving_time} shows the cloud mass evolution from the multiband irradiating flux runs. TLMF\_10 has the same UV flux as TLUV (black), and same IR flux as TLIR\_H (red). Despite the presence of extra IR radiation flux that is an order of magnitude higher than UV flux, the cold gas mass evolution is similar to TLUV. Such behaviour also reflect the fact that the overheating of cold gas is dominated by mixing with hot background medium, which is triggered by the cloud re-expansion after interacting with the UV radiation. Since the UV fluxes are the same, the surviving timescale is similar. However, the additional IR radiation provides a larger acceleration and the cloud in TLMF\_10 reaches a higher final velocity. By the time we stopped the simulation, TLMF\_10 is accelerated to $\sim 203.4$km/s and moved $20.1$pc from start position, about $50\%$ of cold gas in the cloud has been heated over $1500$K. 

Lowering the UV flux fraction in the TLMF\_5 and TLMF\_1 runs produces mass evolution increasingly similar to the pure IR irradiation case. Note the difference in mass and average density (Figure~\ref{fig:2freq_avg}) evolution between $5\%$ and $1\%$ of UV radiation flux is somewhat sharp. Comparing the different UV fraction runs in Figure~\ref{fig:2freq_vvdispmass}, shows that increasing UV radiation from $1\%$ to $5\%$ does not significantly increase bulk acceleration because of the small absolute value of UV flux. But the effect on the compression of the cloud is much more significant.  Hence, even a modest UV fraction can disrupt the cloud without significantly improving the acceleration.  However, the similarity between the TLUV and TLMF\_10 runs also suggests it is not simply the UV fraction that matters since these runs have UV fractions of 100\% and 9\% of the total (IR and UV) flux, respectively.  Our results suggest that if the UV flux and optical depth are large enough to produce significant differential acceleration, compression of the cloud will drive re-expansion and disrupt the cloud in a manner that will significantly enhance mixing with the hot background flow.  This would suggest that the most optically thick star-forming environments, such as ULIRGs where the vast majority of stellar light is reprocessed into the infrared, may be the most efficient locations for driving molecular outflows if radiation pressure dominates.

\subsection{Model Uncertainties and Approximations}

The primary goal of this work is to examine the relative role of UV and IR radiation pressure in accelerating outflows of cold molecular gas that are observed in star-forming galaxies.  Since we focus on this mechanism, we implicitly ignore other possibilities such as entrainment in hot outflows or acceleration due to cosmic ray pressure \citep[e.g.][]{2015ApJ...805..158S,2015MNRAS.449....2M,2018ApJ...854..110Z,2019arXiv190301471W}.  In principle, these other acceleration mechanism may all act in concert to drive outflows or radiation may be an entirely subdominant process.  Our primary motivation for neglecting other acceleration mechanisms is that it allows us to focus on and better understand the radiation hydrodynamics, but given the uncertainties, we view radiation dominated acceleration as a physically plausible limit.

Due to the expense of solving the radiation transfer equations, even the mostly 2D simulations presented here are relatively computationally expensive.  This requires us to make trade-offs in our modeling.  Our simulations focus on the radiation hydrodynamics of the acceleration process, but improve on earlier work by studying the interaction of dusty gas with a multiband irradiating flux under differing assumptions about the optical depths and UV and IR fluxes.  However, we simplify or neglect some of the more complex physics that may be relevant to realistic outflows.  Future work could benefit from including the overall galaxy gravitational potential and any self-gravity of the gas, studying the effects of magnetic fields \citep{2015MNRAS.449....2M}, modeling dilution of the radiation field far from the original source, including additional complexity in the background interstellar medium, a detailed treatment of photoionization \citep[e.g][]{2014MNRAS.443.2018N}, modeling of conduction \citep{2016ApJ...822...31B}, or the process of cloud formation \citep[e.g.][]{2015ApJ...804..137P,2016MNRAS.460L..79W}. We also utilized a simplified prescription for the dust opacity \citep{2003A&A...410..611S} and adopted simple temperature criterion to determine when the dust is destroyed or decoupled \citep{2013MNRAS.434.2329K} from the gas. Future work may benefit from more elaborate treatments of the dust, including its coupling to gas, destruction mechanisms, and opacity.

\section{Conclusion}\label{sec:conclusion}

We consider the effect of UV radiation pressure acceleration of cold, dusty gas.  In contrast to earlier work that focused on IR radiation alone, we find that replacing the IR with UV radiation or including a large fraction of UV radiation is generally detrimental to the cloud survival.  This is due to the UV radiation pressure distorting and compressing the cloud, driving mixing with the hotter background ISM, with mixing ultimately leading to overheating and dust destruction.  In contrast, simulations dominated by IR radiation are more robust, with longer survival times in agreement with earlier work \citep{2018ApJ...854..110Z}.  We attribute this difference to the IR radiation's role in both accelerating the cloud but also in providing an internal radiation pressure due to dust reemission that maintains a more uniform cloud structure and limits mixing.

We also consider the impact of optical depth on the cloud dynamics.  All simulations considered here are optically thin to the IR, but range from optically thin ($\tau_{\rm uv} = 0.13$) to optically thick ($\tau_{\rm uv} = 13$) in the UV.  Generally speaking, decreases in the UV optical depth of the cloud lead to faster disruption times. For moderate to large optical depths, this happens after an initial phase of compression and re-expansion but for the optically thin runs, the process is nearly continuous with no overall compression of the cloud.

With the high radiation fluxes considered here, the UV driven cloud can be accelerated to reasonably high velocities ($\gtrsim 100$ km/s) in the relatively short time ($\lesssim 10^5$ yr) they survive, but they do not travel very far from there in initial location with a ``flying distance'' of only a few parsecs.  In contrast, the IR clouds reach the same distances and nearly as large of velocities, but with most of the initial gas still intact after $10^5$ yr.  Hence, we conclude that a radiation field dominated by emission at IR wavelengths is the most optimal for radiation pressure acceleration.  This suggests that radiation pressure acceleration will be most relevant in highly obscured star-forming galaxies where the UV fractions are low.  In contrast, disruption and mixing will likely tend to destroy clouds in more UV dominated starburst galaxies.  These considerations combined with the need for large radiation fluxes \citep{2018ApJ...854..110Z} suggest ULIRGs and high redshift star-forming galaxies as the environments where radiation pressure is most likely to play a role in driving outflows of cold molecular gas.\\

\acknowledgments{
We thank the referee for a helpful and detailed referee report that significantly improved this work. We also thank Evan Scannapieco, Justin Spilker, and Eve Ostriker for helpful conversations as well as Yan-Fei Jiang and Jim Stone for their contributions to the Athena++ radiation module used in this work. This work used the computational resources provided by the Advanced Research Computing Services (ARCS) at the University of Virginia. We also used the Extreme Science and Engineering Discovery Environment (XSEDE), which is supported by National Science Foundation (NSF) grant No. ACI-1053575. This work was supported by the NSF under grant AST-1616171.
}

\vspace{5mm}
\facilities{Rivanna, XSEDE}

\software{Athena++, VisIt,}

\bibliography{ref}

\end{CJK*}
\end{document}